\documentclass[aps,prd,onecolumn,groupedaddress,superscriptaddress,nofootinbib,longbibliography,10pt]{revtex4-2} 
\usepackage{graphics}
\usepackage[mathcal]{euscript}
\usepackage[latin1]{inputenc}
\usepackage{latexsym}
\usepackage{hyperref}
\usepackage{amsmath}    
\usepackage{graphicx}   
\usepackage{verbatim}  
\usepackage{subfigure}  
\usepackage{hyperref}   
\usepackage{bm}
\usepackage{graphicx}
\usepackage{amssymb}
\usepackage{bbm}
\usepackage{float}
\usepackage{slashed}
\usepackage{amsfonts}
\usepackage{euscript}
\usepackage{amsmath}    
\usepackage{mathrsfs} 
\usepackage{bbding}
\usepackage{soul}

\usepackage[usenames,dvipsnames]{xcolor}
\hypersetup{colorlinks=true, citecolor=MidnightBlue, linkcolor=MidnightBlue, urlcolor=MidnightBlue}

\usepackage{pifont}
\usepackage{cancel}
\usepackage[normalem]{ulem}
\usepackage{calrsfs}
\DeclareMathAlphabet{\pazocal}{OMS}{zplm}{m}{n}

\begin{document}%

\title{Geodesically complete black holes}

\author{Ra\'ul Carballo-Rubio}
\email{raul.carballorubio@sissa.it}
\affiliation{SISSA - International School for Advanced Studies, Via Bonomea 265, 34136 Trieste, Italy}
\affiliation{IFPU - Institute for Fundamental Physics of the Universe, Via Beirut 2, 34014 Trieste, Italy}
\affiliation{INFN Sezione di Trieste, Via Valerio 2, 34127 Trieste, Italy}
\author{Francesco Di Filippo}
\email{francesco.difilippo@sissa.it}
\affiliation{SISSA - International School for Advanced Studies, Via Bonomea 265, 34136 Trieste, Italy}
\affiliation{IFPU - Institute for Fundamental Physics of the Universe, Via Beirut 2, 34014 Trieste, Italy}
\affiliation{INFN Sezione di Trieste, Via Valerio 2, 34127 Trieste, Italy}
\author{Stefano Liberati}
\email{liberati@sissa.it}
\affiliation{SISSA - International School for Advanced Studies, Via Bonomea 265, 34136 Trieste, Italy}
\affiliation{IFPU - Institute for Fundamental Physics of the Universe, Via Beirut 2, 34014 Trieste, Italy}
\affiliation{INFN Sezione di Trieste, Via Valerio 2, 34127 Trieste, Italy}
\author{Matt Visser}
\email{matt.visser@sms.vuw.ac.nz}
\affiliation{School of Mathematics and Statistics, Victoria University of Wellington; PO Box 600, Wellington 6140, New Zealand}

\begin{abstract}
The 1965 Penrose singularity theorem demonstrates the utterly inevitable and unavoidable formation of spacetime singularities under physically reasonable assumptions, and it remains one of the main results in our understanding of black holes. It is standard lore that quantum gravitational effects will always tame these singularities in black hole interiors. However, the Penrose's theorem provides no clue as to the possible (non-singular) geometries that may be realized in theories beyond general relativity as the result of singularity regularization. In this paper we analyze this problem in spherically symmetric situations, being completely general otherwise, in particular regarding the dynamics of the gravitational and matter fields. Our main result is that, contrary to what one might expect, the set of regular geometries that arises is remarkably limited. We rederive geometries that have been analyzed before, but also uncover some new possibilities. Moreover, the complete catalogue of possibilities that we obtain allows us to draw the novel conclusion that there is a clear tradeoff between internal and external consistency: One has to choose between models that display internal inconsistencies, or models that include significant deviations with respect to general relativity, which should therefore be amenable to observational tests via multi-messenger astrophysics.
\end{abstract}

\maketitle
\hrule
\smallskip
\tableofcontents
\bigskip
\hrule
\clearpage
\section{Introduction \label{sec:intro}}
\def\defocus{{\mathrm{defocus}}}

Black holes represent very simple and elegant solutions of the Einstein field equations. It was initially believed that such solutions were a mere mathematical curiosity without a physical counterpart, as it was conjectured that dynamical evolution would have prevented such objects from being formed. Singularity theorems completely reversed this prejudice, proving that the dynamical laws encoded in general relativity do unavoidably lead to the formation of singularities \cite{Senovilla2018}. The first result along this line is due to Penrose \cite{Penrose1964}. The original Penrose singularity theorem assumes the weak energy condition and global hyperbolicity. In 1970 Hawking and Penrose \cite{Hawking:1970} showed that it is possible to get rid of the hypotheses of global hyperbolicity by imposing the strong energy condition instead.

Therefore, if we restrict our attention to general relativity, black holes represent (possibly together with naked singularities if the cosmic censorship is violated \cite{Penrose:1969}) the only possible outcome of many physically realistic gravitational collapses. However, it is reasonable to assume that, once that quantum gravity effect are taken into account, singularities will be regularized. The structure and properties of the corresponding non-singular black holes is largely unknown, in part due to our lack of knowledge about the dynamics in quantum gravity, although there have been many and diverse attempts at constructing geometries or toy models that aim at capturing the leading effects beyond general relativity \cite{Bardeen1968,Hayward2006,Frolov2014,Carballo-Rubio2018,Mathur:2005zp,Brustein2018,Brustein2019,Perez2017,Ashtekar:2018cay,Ashtekar:2018lag,Buoninfante2018,Buoninfante2019}.  

In this paper, we take a completely different approach, remaining agnostic to the particular quantization scheme of the gravitational field and using Penrose's theorem as guidance. We shall build upon Penrose's theorem, which tells us that singularities are inevitable in general relativity under specific conditions imposed on matter fields, and perform a systematic analysis of the non-singular possibilities that is blind to the dynamics (classical, semi-classical, or quantum) and assumes a minimal set of kinematical requirements. Our analysis is quite general, applying to any theory of quantum gravity that satisfies these kinematical requirements in a certain regime, and virtually any theory of modified gravity; however, it is important to keep in mind that in this first paper we work in spherically symmetric situations for simplicity. 

Moreover, we decide to follow the logic of the original Penrose theorem rather than the logic of the subsequent Hawking and Penrose theorem, even if the former has the extra hypothesis of global hyperbolicity, because the structure of the original Penrose theorem suits our purposes better. The reason is that it is possible to schematically divide Penrose theorem into two parts. In the first part of the theorem, the weak energy condition and the Einstein field equations are used in order to prove the existence of a focusing point. Then, considerations of a purely topological nature are used in order to prove that the existence of a focusing point is incompatible with geodesic completeness. Hence, in order to understand the possible geometries that follow from the violation of either the weak energy condition or the Einstein field equations, we just need to categorize in geometric terms the possible ways in which the focusing point can be avoided.

The paper is organized as follows. In section \ref{sec:penrth} we briefly review the original Penrose theorem and explain in depth the rationale of our analysis. In section \ref{sec:geom} we explain the geometrical setting that we will be using, and prove some results that will be important for the subsequent discussion. Our main thesis is formalized in section \ref{sec:cases}, where we categorize and discuss the possible geometries. While in this paper we focus on the geometrical analysis, we briefly discuss some of its physical implications in section \ref{sec:phen}, which are investigated in more detail in a companion letter \cite{letter}. Finally, section \ref{sec:discussion} contains a discussion of our main conclusions.

\section{Avoiding singularities: Beyond Penrose's theorem \label{sec:penrth}}

Penrose's singularity theorem \cite{Penrose1964}, and its modern variants and extensions (see for instance \cite{Senovilla2014}), demonstrates that
once a closed trapped surface $\mathscr{S}^2$ is formed, then in some region of spacetime contained in the causal future $J(\mathscr{S}^2)$ of $\mathscr{S}^2$, one of the following must hold in order to avoid a spacetime singularity (more specifically, to avoid that the spacetime is geodesically incomplete):
\begin{itemize}
\item[(a)]{The weak energy condition is violated.}
\item[(b)]{The Einstein field equations do not hold.}
\item[(c)]{Pseudo-Riemannian geometry does not provide an adequate description of spacetime.}
\item[(d)]{Global hyperbolicity breaks down.}
\end{itemize}
These options span a huge space of possibilities that are very difficult (if not impossible) to describe exhaustively. Aside from the different possible combinations of these possibilities one can devise, each of them independently can also lead to numerous diverse scenarios. 

\begin{figure}[!h]%
\begin{center}
\vbox{\includegraphics[width=0.25\textwidth]{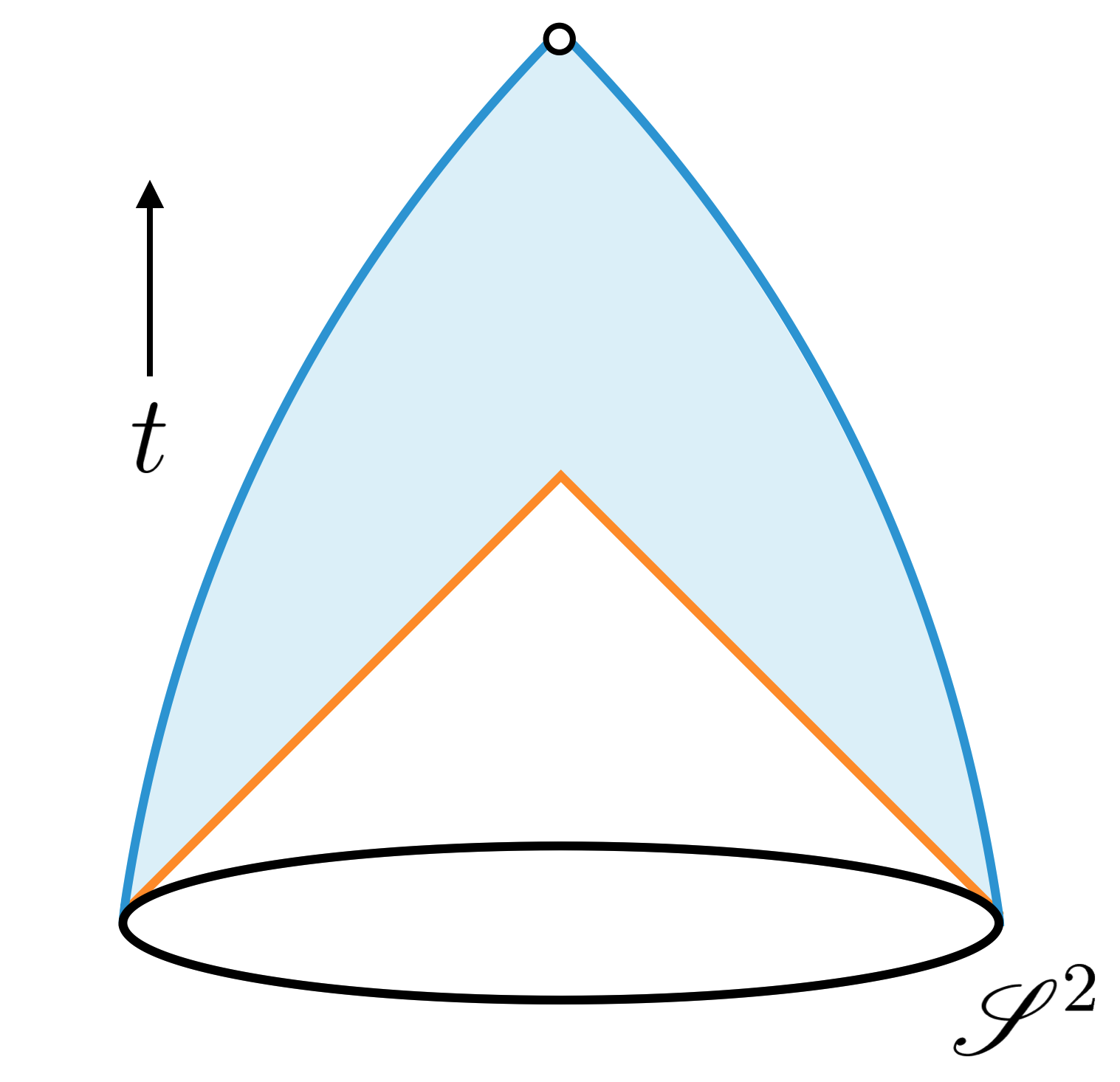}}
\bigskip%
\caption{In a nutshell, Penrose's theorem implies that there cannot exist a non-pathological focusing point to the future of the assumed trapped surface. Otherwise the boundary of the causal future of the trapped surface $\mathscr{S}^2$ would be compact, which leads to a contradiction with the non-compact nature of an initial Cauchy surface $\mathscr{C}^3$ \cite{Penrose1964}.}
\label{fig:fig1}%
\end{center}
\end{figure}%

Not all these different possibilities (a)--(d) carry the same weight from a physical perspective. In particular, the first two possibilities, (a) and (b), are widely expected to describe features of any theory that brings the principles of quantum mechanics into the mix (such as semiclassical gravity), modifies the dynamics of the gravitational field, or does both (as in any quantum gravity approach). The status of the remaining two possibilities, (c) and (d), is more subtle. It is reasonable to think that, in a theory of quantum gravity, spacetime may lose its smoothness in certain situations, so that a description in terms of differentiable manifolds breaks down. If this happens in a bounded region, it seems to be possible quite generically to devise a classical geometry that matches the physical one at the boundary of this non-classical region; while one does not expect this effective classical description to provide a precise description of the non-classical regime, it may capture some of the relevant physics of the transition through this region. 

Even more importantly, advances in frameworks such as loop quantum gravity and loop quantum cosmology show that a description in terms of an effective metric provides a good approximation in many situations \cite{Kaminski2010,Ashtekar2011,Ashtekar2015}, so that, for a large sector of initial conditions for the quantum state of the system, the description in terms of differentiable manifolds remains meaningful throughout dynamical evolution. Whether these initial conditions are relevant to describe physical situations of interest in black holes (where several technical issues remain to be solved \cite{Olmedo2016,Campiglia2016,Alesci2018}) is unclear at the moment but, regardless of this issue, we think that there is no question that effective descriptions in terms of differentiable manifolds satisfying (c) and (d) above can provide useful complementary illustrations of the physics at play.
Moreover, one cannot expect that quantum gravity will simply regularize all singularities by breaking the description in terms of differentiable manifolds without modifying the spacetime anywhere else, as this would imply that the gravitational Hamiltonian is not bounded from below \cite{Horowitz1995}. Until a definite rule that can be used to determine precisely which singularities can safely be regularized using quantum effects, we think that addressing the regularization of singularities within the quite general formalism of smooth manifolds is an interesting problem.
With this idea in mind, our goal is thus gaining a complete understanding of the information that pseudo-Riemannian geometry may contain about theories that go beyond general relativity, and regularize the singularities of the latter. Accordingly, we will work in a framework where we \emph{ab initio} impose the minimal conditions that:
\begin{itemize}
\item[(1)]{Pseudo-Riemannian geometry provides an effective description of spacetime.}
\item[(2)]{The spacetime is globally hyperbolic.\footnote{Note that Penrose's theorem also assumes that the Cauchy surface is non-compact. As explained in \cite{Hawking1973} the non-compact assumption can be relaxed by assuming the existence of at least one future inextendible curve from the Cauchy surface $\mathcal{C}^3$ which does not intersect the causal future of the trapping surface. We implicitly assume that this is realized, as it corresponds to the very physically reasonable assumption that there exists at least one observer which does not fall into the collapsing star.}}
\item[(3)]{The spacetime is geodesically complete.}
\item[(4)]{There are no curvature singularities.\footnote{It is worth mentioning the existence of works in the literature in which assumption (2) holds but this one is dropped, such as \cite{Olmo2015,Olmo2016,Bejarano2017,Olmo2017}.}}
\end{itemize}
On the other hand, we will accept that both (a) and (b) in Penrose's list of possibilities can take place, perhaps simultaneously. We will see that the geometric structure assumed in (1), (2), (3) and (4) above is tight enough to leave only a handful of possibilities.

Some of these assumptions have been widely used in previous explorations of singularity regularization in black holes \cite{Bardeen1968,Frolov1979,Frolov1981,Roman1983,Dymnikova1992,Borde1996,Hayward2006,Bronnikov2006,Dymnikova2001,Ansoldi2008,Frolov2014,Frolov2016,Carballo-Rubio2018}; however, it is worth remarking that (strict) global hyperbolicity is violated in static regular black holes (we will discuss this explicitly in due course). What makes our approach novel, and more powerful, is that all these works assume additional postulates in order to select specific realizations of non-singular spacetimes, while here we want to provide a complete catalogue of possibilities based only on geometric notions. Our study can be understood as a supplement to Penrose's theorem that goes a step beyond and characterizes all the spherically symmetric geometries that avoid the formation of focusing points while satisfying (1--4) above.

\section{Geometric setting}\label{sec:geom}

Let us start with some geometric preliminaries. We will be dealing with a 4-dimensional and globally hyperbolic spacetime $\mathscr{M}$, with $t:\mathscr{M}\rightarrow\mathbb{R}$ a global time function and $\mathscr{C}^3$ a Cauchy surface\footnote{Let us stress that we refer to global hyperbolicity in the ``strict" sense of existence of a Cauchy hypersurface which is intersected by every inextensible causal curve exactly once. The weaker condition of partial global hyperbolicity \cite{Hawking1973}, in which there is a partial Cauchy hypersurface which is intersected by every inextensible, causal curve at most once, has also been used in explorations of black hole spacetimes \cite{Simpson2018,Simpson2019}.}. Additionally, our spacetimes of interest describe the collapse of a regular distribution of matter from a given initial Cauchy surface with topology $\mathbb{R}^3$. It is important to note that this fixes the topology of the entire manifold as our assumptions are incompatible with topology change \cite{Geroch1967,Geroch1970,Borde1994}. It is also useful to restrict the initial discussion to spherically symmetric spacetimes, postponing for future works the discussion of the more general setting in which this assumption is dropped. Let us recall the standard property that the isometry group of a spherically symmetric spacetime permits to identify a foliation in 2-spheres $S^2$. Furthermore, spherical symmetry allows us to prove the following auxiliary result:

\vspace{0.2cm}

\noindent
\textbf{Proposition 1:} Under the assumptions (1--4) above, the entire manifold $\mathscr{M}$ can be covered with a single coordinate chart in which all points $p\in\mathscr{M}$ are labelled using double null coordinates (associated with radial null geodesics) and two angular coordinates.\footnote{Even though this result is certainly not new, there is no proof of this proposition in the literature that we are aware of; we have included a proof for completeness.}

\noindent
\textbf{Proof:} Due to the existence of a local light-cone structure and the dimensionality of $\mathscr{M}$, through every point $p$ in the manifold there are exactly two future-directed radial null geodesics (which we will call in the following left-going and right-going) and two past-directed radial null geodesics (their continuation to the past of $p$). Global hyperbolicity implies that either the two future-directed or the two past-directed causal curves through $p$ must intersect $\mathscr{C}^3$. As there are only two such geodesics for each $p$, we can use these (together with the two angular coordinates on the unit 2-sphere) in order to label all points in spacetime, as long as these labels are unique. In order to show that these labels are unique, let us demonstrate that any situation in which these are not unique is inconsistent with our assumptions. For two points $p,q\in\mathscr{M}$, with $p< q$, let us define their causal diamond $\mathscr{D}(q,p)$ as the intersection between the causal future of $p$ and the causal past of $q$, namely $\mathscr{D}(q,p)=J^-(q)\cap J^+(p)$ (for fixed values of the angular coordinates). If the two radial null geodesics crossing at $p$ and intersecting $\mathscr{C}^3$ also cross at $q$, the boundary of $\mathscr{D}(q,p)$ is smooth except at the points $p$ and $q$; in other words, the sides of the diamond are smoothed, as each of them are identified with segments of a left-going and right-going radial null geodesic, respectively. This smoothing implies the existence of at least two points in the boundary of $\mathscr{D}(q,p)$ in which the light cones in the corresponding tangent spaces are degenerate, as $\Sigma^2$ is simply connected. Let us consider for instance the side of the diamond generated by the left-going radial null geodesic; it is clear that there exist right-going null geodesics that cross the former at two points, but also other right-going null geodesics without crossing points (this is also true for smaller segments of the left-going radial null geodesic). Smoothness implies the existence of an intermediate right-going radial null geodesic that is tangent to the segment of the corresponding left-going radial null geodesic at one point at least; the light cone at this point would be then degenerate. $\blacksquare$

\begin{figure}[!h]%
\begin{center}
\vbox{\includegraphics[width=0.8\textwidth]{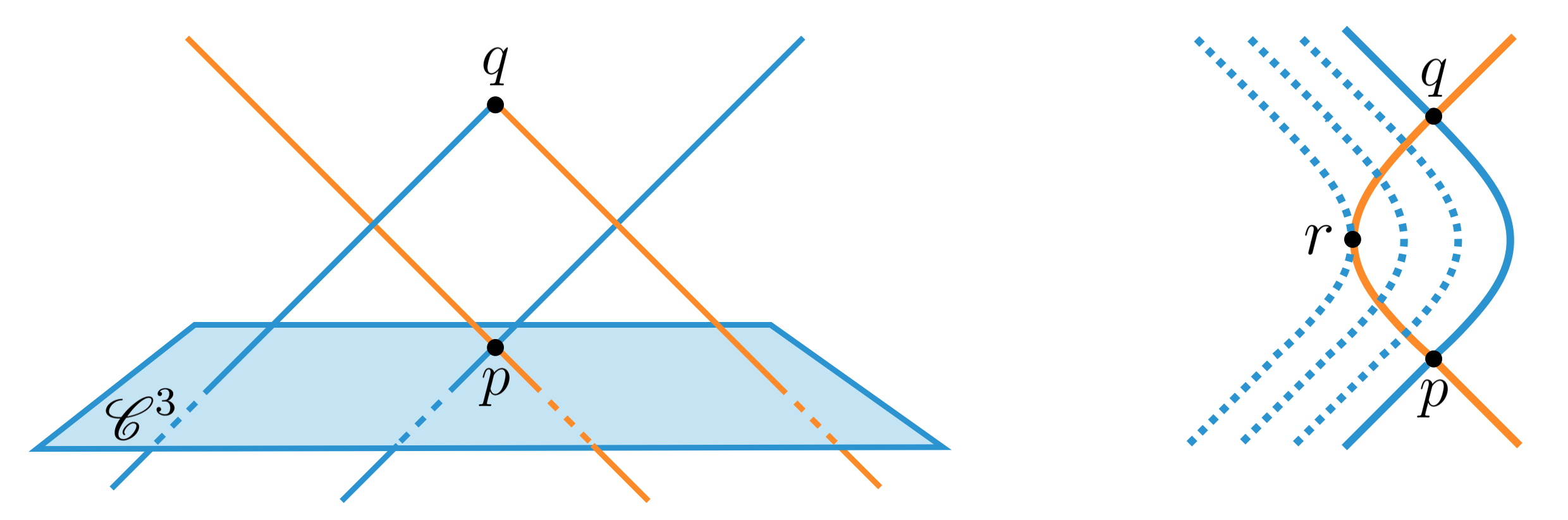}}
\bigskip%
\caption{\emph{On the left:} Left-going and right-going radial null geodesics that can be used to label two points $p$ and $q$ in spacetime, and the corresponding causal diamond. \emph{On the right:} The curved lines between $p$ and $q$ illustrate the hypothetical situation discussed in \textbf{Proposition 1}, where the sides of the causal diamond are smoothed. The dashed lines mark some additional radial null geodesics, illustrating that the light-cone structure of spacetime becomes degenerate on two points along these sides (for graphical simplicity we are just including one of these points, $r\in\mathscr{M}$).}
\end{center}
\end{figure}%

The areas of the 2-spheres preserved by rotations are geometric invariants which, in turn, provide a suitable definition for the radial coordinate $r$ in these spacetimes. We will often use the coordinates $(v,r)$ and $(u,r)$ (in the following, we omit the angular coordinates) instead of $(u,v)$. For instance, in the former case we always write the line element, without loss of generality, as
\begin{equation}\label{eq:linel1}
\text{d}s^2=-e^{-2\phi(v,x)}F(v,x)\text{d}v^2+2e^{-\phi(v,x)}\text{d}v\text{d}x+r^2(v,x)\text{d}\Omega^2,
\end{equation}
where $\text{d}\Omega^2$ is the line element on the unit 2-sphere. The function $r(v,x)$ may be, for $v$ fixed, bijective. In this case, we can perform a change of coordinates in order to write 
\begin{equation}\label{eq:linel}
\text{d}s^2=-e^{-2\phi(v,r)}F(v,r)\text{d}v^2+2e^{-\phi(v,r)}\text{d}v\text{d}r+r^2\text{d}\Omega^2.
\end{equation}
Even if $r(v,x)$ for $v$ fixed does not provide a bijective relation between $r$ and $x$, we can always find open subsets in the domain of $r(v,x)$ in which a bijective relation exists. Hence, we would at worst have different coordinate charts in each of which the line element has the form given above in Eq. \eqref{eq:linel}, covering different regions of spacetime.

As in Penrose's theorem, we will assume the existence of a spacelike trapped surface $\mathscr{S}^2$, defined using the null vector fields that are normal to it. Due to the dimensionality of $\mathscr{M}$ and $\mathscr{S}^2$, and the spacelike character of the latter, there are two independent (future-directed) normal null vectors at each point of $\mathscr{S}^2$, that we will call $\bm{l}$ (outgoing null normal) and $\bm{k}$ (ingoing null normal). If $h_{ab}$ is the 2-metric induced  on $\mathscr{S}^2$, we can define the expansions along these vector fields as
\begin{equation}
\theta^{(\bm{X})}={1\over\sqrt{h}} \; \mathcal{L}_{\bm{X}}\sqrt{h}=h^{ab}\nabla_a X_b,\qquad \bm{X}\in\{\bm{l},\bm{k}\},
\end{equation}
where $\mathcal{L}_{\bm{X}}$ is the Lie derivative along $\bm{X}$, $h=\mbox{det}(h_{ab}$) and $\bm{\nabla}$ is the 4-dimensional covariant derivative. The expansion $\theta^{(\bm{X})}$ measures the local change in the area of $\mathscr{S}^2$ under a local deformation along the vector field $\bm{X}$. Then, a trapped surface is defined by the conditions
\begin{equation}
\theta^{(\bm{k})}<0,\qquad\qquad \theta^{(\bm{l})}<0.
\end{equation}

This definition connects naturally with quasi-local characterizations of the boundary of black holes. Although for the discussion in this paper we only need the existence of a trapped surface, it is worth keeping in mind that the geometries that we will be analyzing display the same quasi-local properties as a black hole only for finite (in some cases very large) periods of time, but without necessarily sharing their global properties, or other features such as the existence of spacetime singularities. The study of these quasi-local properties has been a very active area of research during the last two decades and there are many published works \cite{Hajicek1973,Hayward1993,Ashtekar1998,Ashtekar2002,Ashtekar2003,Ashtekar2005,Booth2003,Hayward2004,Hayward2006b,Andersson2005,Andersson2007}, as well as several reviews  \cite{Ashtekar2004,Krishnan2007,Gourgoulhon2008} available. One of these quasi-local concepts is that of a \emph{future outer trapping horizon} \cite{Hayward1993}: a 3-dimensional hypersurface $\mathscr{H}=\bigcup_{t\in\mathbb{R}}\mathscr{S}_t$ of $\mathscr{M}$ that is foliated by closed and spacelike 2-dimensional surfaces $\mathscr{S}_t$ satisfying $\theta^{(\bm{k})}<0$ and $\theta^{(\bm{l})}=0$, which can be understood as the limiting (or marginal) case of a trapped surface. There is the additional condition $\mathscr{L}_{\bm{k}}\theta^{(\bm{l})}<0$ which guarantees that the spacetime around the horizon $\mathscr{H}$ has the quasi-local structure associated with the outer horizon of a black hole \cite{Hayward1993}, in contradistinction to the inner horizon that appears in charged and rotating black holes. In particular, this third condition implies the existence of trapped surfaces arbitrarily close (from inside) to the marginally trapped surface which, again, is the main ingredient that we need for our discussion.


\section{Taxonomy of non-singular geometries \label{sec:cases}}

We are now almost in position to formulate our central thesis, but before doing so, we still need to demonstrate another auxiliary result:

\vspace{0.2cm}

\noindent
\textbf{Proposition 2:} If a congruence of outgoing radial null geodesics has a focusing point at a finite affine distance $\lambda=\lambda_0$, then $\left.\theta^{(\bm{l})}\right|_{\lambda=\lambda_0}=-\infty$.

\noindent
\textbf{Proof:} Let us recall that the expansion $\theta^{(\bm{l})}$ measures the rate of change of the element of area that is normal to the congruence of outgoing null geodesics \cite{Poisson2009}, namely
\begin{equation}
\theta^{(\bm{l})}=\frac{1}{\delta A^{(\bm{l})}}\;\frac{\text{d}}{\text{d}\lambda}\delta A^{(\bm{l})},
\end{equation}
where $\delta A^{(\bm{l})}$ is the cross-sectional area of the congruence and $\lambda$ the affine parameter along the congruence. If we choose the available freedom in the definition of the affine parameter so that $\lambda=0$ corresponds to $\mathscr{S}^2$, we then have
\begin{equation}\label{eq:daeq}
\ln\left(\frac{\left.\delta A^{(\bm{l})}\right|_{\lambda=\lambda_0}}{\left.\delta A^{(\bm{l})}\right|_{\lambda=0}}\right)=\int^{\lambda_0}_0\text{d}\lambda\,\theta^{(\bm{l})}(\lambda).
\end{equation}
By construction, the quantity $\left.\delta A^{(\bm{l})}\right|_{\lambda=0}$ is finite and positive. If there is a focusing point at $\lambda=\lambda_0$, $\left.\delta A^{(\bm{l})}\right|_{\lambda=\lambda_0}=0$, and the logarithm goes to infinitely negative values. Using the bound
\begin{equation}
\left|\int^{\lambda_0}_0\text{d}\lambda\,\theta^{(\bm{l})}(\lambda)\right|\leq \lambda_0\sup_{\lambda\in [0,\lambda_0]} \left|\theta^{(\bm{l})}(\lambda)\right|,
\end{equation}
it follows that the expansion $\theta^{(\bm{l})}(\lambda_0)$ must be divergent (and negative). $\blacksquare$

\vspace{0.2cm}

\begin{figure}[!h]%
\begin{center}
\vbox{\includegraphics[width=0.9\textwidth]{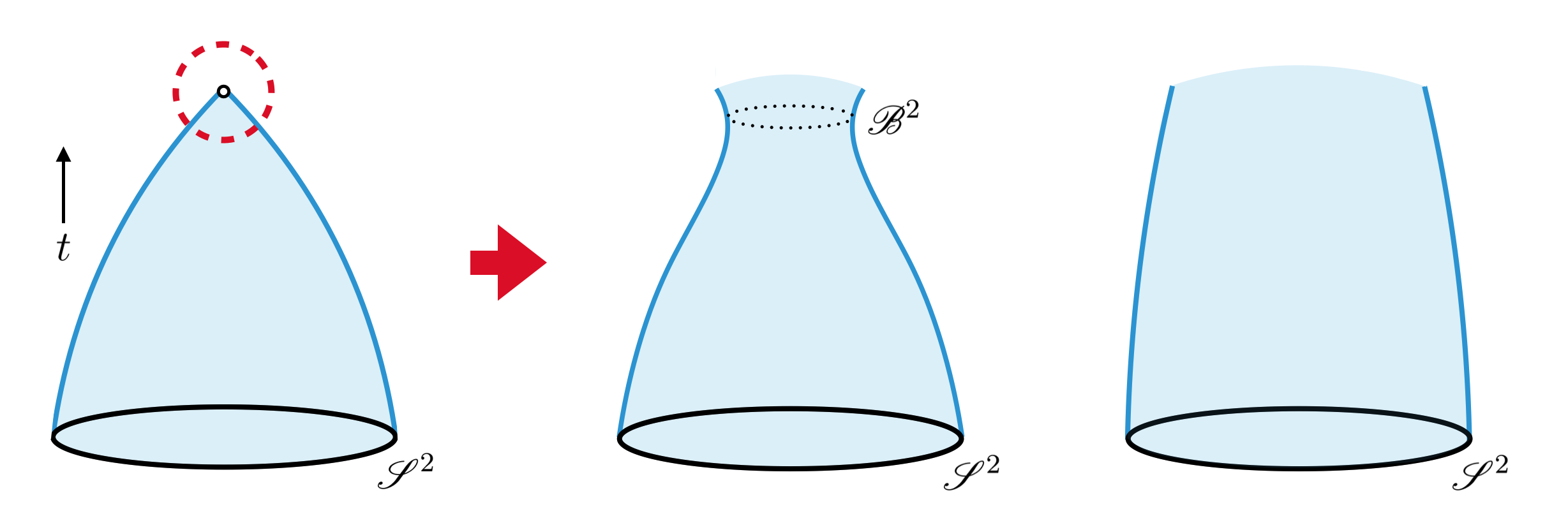}}
\bigskip%
\caption{In order to avoid that the spacetime is geodesically incomplete, the spacetime geometry must be modified in the surroundings of the focusing point in Penrose's theorem, so that either a defocusing point is created at a finite affine distance (thus also creating the 2-surface $\mathscr{B}^2$) or infinite affine distance, or the focusing point is displaced to infinite affine distance. The figure in the right is compatible with the two last cases. Note that ingoing radial null geodesics are not included in this picture, as these can display different behaviors that are analyzed in detail in the text.}
\label{fig:fig2}%
\end{center}
\end{figure}

The converse statement is also true but less immediate to prove. Indeed, if the expansion $\theta^{(\bm{l})}$ is divergent and negative, then the transverse area $A^{(\bm{l})}$ becomes zero, leading to focusing points; a proof of this statement can be found in \cite[Sec. 9.3]{Wald1984}. Thus, following Penrose's theorem and under conditions (1), (2), and (3) as defined in Sec. \ref{sec:penrth}, in order to avoid that the spacetime is geodesically incomplete we need to modify the spacetime geometry in the vicinity of the focusing point, either creating a defocusing point or displacing the focusing point to infinite affine distance (see Fig. \ref{fig:fig2}). The expansion $\theta^{(\bm{l)}}(\lambda)$ must then remain finite for all the possible values of $\lambda\in[0,\infty)$ (where we are again identifying again $\lambda=0$ with $\mathscr{S}^2$ without loss of generality).

The limiting situation in which the focusing point is displaced to infinite affine distance will be shown later to be still singular, although due to curvature invariants blowing up instead of geodesic incompleteness, thus violating our condition (4) in Sec. \ref{sec:penrth}. Hence, we can anticipate that, in practice, the most interesting geometries are those with a defocusing point at $\lambda=\lambda_\defocus$, where $\lambda_\defocus$ can be either finite or infinite. In terms of the behavior of $\theta^{(\bm{l)}}(\lambda)$ we can distinguish three possibilities, as the outgoing expansion can either remain negative but finite (thus having no defocusing points), vanish asymptotically for infinite affine distance (whether or not there are defocusing points depends on the convergence properties of the integral of the expansion), or vanish at a finite affine distance (thus having a defocusing point at a finite affine distance). These three possibilities can be further characterized taking into account the behavior of congruences of ingoing null geodesics that intersect the outgoing ones for these values of $\lambda$ and, in particular, their expansion $\theta^{(\bm{k})}$ at the intersection points. The intersection between ingoing and outgoing null geodesics will take place at a radius $r=R_\defocus$, so we will denote the corresponding value of the ingoing expansion by $\bar{\theta}:= \left.\theta^{(\bm{k})}\right|_{r=R_\defocus}$. It is then convenient to use a label with three entries $\left(\lambda_\defocus,R_\defocus,\bar{\theta}\gtreqless0\right)$ in order to characterize these possibilities, with the first entry being the value of the affine parameter for which $\theta^{(\bm{l)}}(\lambda)$ vanishes, the second entry being the value of the radius where the expansion vanishes or, if the latter does not vanish, its $\lambda\rightarrow\infty$ limit, and the third entry being the sign of the expansion $\theta^{(\bm{k})}$ of the ingoing null geodesic that intersects the outgoing one at the very same value of the radius. 

Using these labels, there are 8 possibilities: $(\lambda_0,R_0,\bar{\theta}<0)$, $(\lambda_0,R_0,\bar{\theta}\geq0)$, $(\infty,R_\infty,\bar{\theta}<0)$, $(\infty,R_\infty,\bar{\theta}\geq0)$, $(\infty,0,\bar{\theta}<0)$, $(\infty,0,\bar{\theta}\geq0)$, $(\emptyset,0,\bar{\theta}<0)$, $(\emptyset,0,\bar{\theta}\geq0)$, where $\emptyset$ means that there is no value of $\lambda$ for which the expansion $\theta^{(\bm{l})}(\lambda)$ vanishes. Let us describe these possibilities in more detail (a flowchart of this classification is provided in Fig. \ref{fig:diagram}):

\begin{description}

\item[Case A -]{Defocusing point at a finite affine distance, $\lambda_\defocus=\lambda_0$:}
\begin{itemize}
\item[A.I:]{$(\lambda_0,R_0,\bar{\theta}<0)$: 
The expansion $\theta^{(\bm{l})}$ vanishes and changes sign at a finite affine distance $\lambda=\lambda_0$ or, in terms of the radius, at a value $R_0>0$ of the radial coordinate along the congruence of outgoing radial null geodesics at $\lambda=\lambda_0$ (namely, the radius of $\mathscr{B}^2$ in Fig. \ref{fig:fig2}). On the other hand, the expansion of the intersecting ingoing radial null geodesics remains negative until (and including) $\lambda_0$, so that $\left.\theta^{(\bm{k})}\right|_{r=R_0}<0$.}

\item[ A.II:]{$(\lambda_0,R_0,\bar{\theta}\geq0)$: 
The only difference with respect to the previous case is that the expansion of the intersecting ingoing radial null geodesics does not remain negative, $\left.\theta^{(\bm{k})}\right|_{r=R_0}\geq0$.}
\end{itemize}

\item[Case B - ]{Defocusing point at an infinite affine distance, $\lambda_\defocus=\infty$:}
\begin{itemize}
\item[B.I:]{$(\infty,R_\infty,\bar{\theta}<0)$: 
The expansion $\theta^{(\bm{l})}$ vanishes in the limit $\lambda\rightarrow\infty$, in a manner such that the integral in Eq. \eqref{eq:daeq} is convergent. The corresponding asymptotic value of the radial coordinate for radial outgoing null geodesics is $R_\infty>0$. The expansion of the intersecting ingoing radial null geodesics remains negative, so that $\left.\theta^{(\bm{k})}\right|_{r=R_\infty}<0$.}
\item[B.II:]{$(\infty,R_\infty,\bar{\theta}\geq0)$: 
The only difference with respect to the previous case is that the expansion of the intersecting ingoing radial null geodesics does not remain negative, $\left.\theta^{(\bm{k})}\right|_{r=R_\infty}\geq0$.}
\item[B.III:]{$(\infty,0,\bar{\theta}<0)$: 
The expansion $\theta^{(\bm{l})}$ vanishes in the limit $\lambda\rightarrow\infty$, in a manner such that the integral in Eq. \eqref{eq:daeq} is divergent. Thus, the radial coordinate vanishes asymptotically along these geodesics (in other words, there is an asymptotic focusing point). The expansion of the intersecting ingoing radial null geodesics remains negative, so that $\left.\theta^{(\bm{k})}\right|_{r=R_\infty}<0$.}
\item[B.IV:]{$(\infty,0,\bar{\theta}\geq 0)$} The only difference with respect to the previous sub-case  is that the expansion of the intersecting ingoing radial null geodesics does not remain negative, $\left.\theta^{(\bm{k})}\right|_{r=R_\infty}\geq0$.
\end{itemize}

\item[Case C - ]{No defocusing point, $\lambda_\defocus=\emptyset$:}
\begin{itemize}
\item[C.I:]{$(\emptyset,0,\bar{\theta}<0)$: 
The expansion $\theta^{(\bm{l})}$ remains negative in the limit $\lambda\rightarrow\infty$, which in particular implies that the integral in Eq. \eqref{eq:daeq} is divergent. Thus, the radial coordinate vanishes asymptotically along these geodesics. The expansion of the intersecting ingoing radial null geodesics remains negative, so that $\left.\theta^{(\bm{k})}\right|_{r=R_\infty}<0$.}

\item[C.II:]{$(\emptyset,0,\bar{\theta}\geq0)$: 
The only difference with respect to the previous sub-case  is that the expansion of the intersecting ingoing radial null geodesics does not remain negative, $\left.\theta^{(\bm{k})}\right|_{r=R_\infty}\geq0$.}
\end{itemize}

\end{description}

It is important to keep in mind that, in practical terms, sub-cases B.III and C.I fall into the same category, as both of these situations can be understood as the limiting case in which the focusing point is pushed to an infinite affine distance along outgoing null geodesics. The same comment applies to cases B.IV and C.II, that only differ from the previous sub-cases (B.III and C.I)  in the behaviour of $\theta^{(\bm{k})}$. Hence, when discussing these cases later, we will do it simultaneously.

In the following five sections we discuss in detail the geometric properties of each of the classes of geometries associated with the different cases defined above describing the behavior of outgoing null geodesics. One must keep in mind that these are not mutually exclusive cases, but that at least one of them must be satisfied (for each outgoing null geodesic crossing the region with trapped surfaces) in order to avoid the manifold being geodesically incomplete or not globally hyperbolic. The continuity of the spacetime manifold implies the existence of other trapped surfaces in an open neighborhood around $\mathscr{S}^2$, and we will consider the simplifying assumption that different outgoing null geodesics stemming from these other trapped surfaces belong to the same case (so that neither of the remaining cases can hold). This will allow us to better understand the geometric implications behind each of these cases; moreover, we can consider the set of geometries arising from this analysis as a kind of minimal complete set, in the sense that any possible geometry can be obtained combining the features found in this set.

\subsection*{Case A.I: Evanescent horizons, $(\lambda_0,R_0,\bar{\theta}<0)$ \label{case:1}}

As indicated above, in analyzing this case we will assume that all outgoing radial null geodesics stemming from trapped surfaces (as well as the intersecting ingoing radial null geodesics) display the same qualitative behavior, so that none of the remaining cases is realized. We will proceed similarly for the analysis of the other cases. In the current case, this implies the existence of a set of points $r=R_0(u)$ where the expansion $\theta^{(\bm{l})}$ vanishes. On the other hand, each ingoing radial null geodesic defines a slice of spacetime that is associated with a constant value of $v$. We can therefore identify the  values of $v$ associated with each $R_0(u)$ through a function $v(R_0(u))$. Let us start with a result constraining the form of this function:

\vspace{0.2cm}

\noindent
\textbf{Proposition 3:} There must exist $2k+1$ (with $k\geq0$) open intervals in the domain of $v(R_0)$ where the latter is bijective and the inverse function $R_0(v)$ exists, which must be either a strictly increasing or strictly decreasing function in $k+1$ of these intervals, and either a strictly decreasing or strictly increasing function in the remaining $k$ intervals.

\noindent
\textbf{Proof:} In spherically symmetric and globally hyperbolic spacetimes, outgoing radial null geodesics cannot cross (otherwise, the light-cone structure of spacetime would be degenerate at the crossing point; see \textbf{Proposition 7} below). In a slice of constant $v$, outgoing radial null geodesics that have a greater value of $r$ have a smaller value of $u$. This implies that the function $R_0(u)$ must be strictly decreasing in order to avoid crossing points between different outgoing radial null geodesics. On the other hand, we show in App. \ref{sec:regconds} that the number of points in a slice of constant $v$ where $\theta^{(\bm{l})}$ vanishes and changes sign must be even in order to ensure the regularity of curvature invariants. If there are only two such points for all values of $v$ (the first one corresponding to the future outer trapping horizon), the function $v(R_0)$ is bijective and we can find its inverse $R_0(v)$. That this function must be either strictly increasing or decreasing follows from the strictly decreasing nature of $R_0(u)$. In the more general situation, there are open intervals in the domain of $v(R_0)$ around the remaining $2k+1$ points where $\theta^{(\bm{l})}$ vanishes where $v(R_0)$ is bijective. $\blacksquare$

\vspace{0.2cm}

Even if the proposition above reduces the number of possibilities, but we can reduce them even further using the regularity conditions derived in Appendix~\ref{sec:regconds}:

\vspace{0.2cm}

\noindent
\textbf{Proposition 4:} The function $R_0(v)$ defined over the innermost open subset of the domain of $v(R_0)$ where the latter is bijective must be strictly increasing in order to avoid curvature singularities at $r=0$. Hence, from the $2k+1$ (with $k\geq0$) open intervals of values of $R_0$ where the inverse function $R_0(v)$ exists, this function must be strictly increasing in $k+1$ and decreasing in $k$ of them.

\noindent
\textbf{Proof:} If $R_0(v)$ defined over the innermost open subset where $v(R_0)$ is bijective is strictly decreasing, the only possibility that is compatible with the assumption that all points in $R_0(v)$ are reached in finite affine distance along some outgoing radial null geodesic is that $R_0(v)$ eventually vanishes. The point in which this function vanishes must be reached in finite affine distance along a specific outgoing radial null geodesic, in order to avoid stepping into another of the cases discussed in Sec. \ref{sec:cases}, namely the case $(\infty,R_\infty,\bar{\theta}<0)$ (this applies both to the particular choice $R_\infty=0$, but also to the situation in which  $\partial R_0(v)/\partial v$ vanishes asymptotically so that $R_0(v)$ tends to a finite value). This implies the existence of a curvature singularity that is reached in finite affine distance along this specific outgoing radial null geodesic, as follows from our arguments in Appendix \ref{sec:regconds}. The argument is straightforward if one takes a slice of constant $v$ that reaches $r=0$ simultaneously with $r=R_0(v)$, and notices that the necessary condition of an even number of zeros of $\theta^{(\bm{l})}$ for $r>0$ is not satisfied. $\blacksquare$

\vspace{0.2cm}

Propositions 3 and 4 imply that there must exist at least a portion of the submanifold in which $\theta^{(\bm{l})}$ vanishes such that $v(R_0)$ is invertible and $R_0(v)$ is strictly increasing. Interestingly, these portions correspond to future inner trapping horizons, as defined by Hayward \cite{Hayward1993}:

\vspace{0.2cm}

\noindent
\textbf{Proposition 5:} Whenever $v(R_0)$ is invertible and $R_0(v)$ strictly increasing, the corresponding subset of the hypersurface in which $\theta^{(\bm{l})}$ vanishes is a spacelike future inner trapping horizon.

\begin{figure}[h]%
\begin{center}
\vbox{\includegraphics[width=0.3\textwidth]{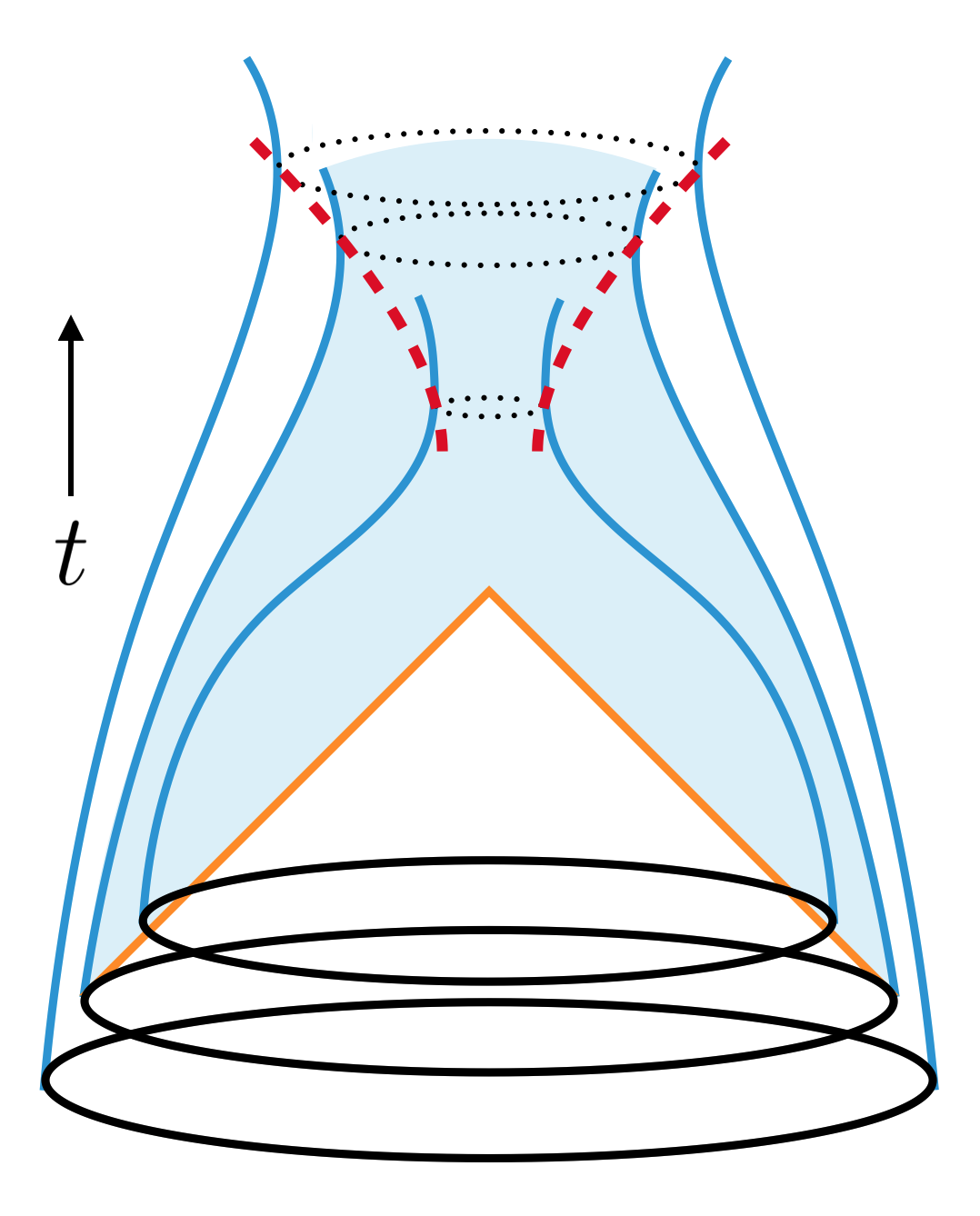}}
\bigskip%
\caption{Determination of the position of the future inner trapping horizon (red dashed line), taking into account different trapped surfaces in the neighborhood of $\mathscr{S}^2$ and tracking the corresponding outgoing null geodesics. In this case the future inner trapping horizon is spacelike.}
\label{fig:fig3}%
\end{center}
\end{figure}%

\noindent
\textbf{Proof:} Following the definition in \cite{Hayward1993}, a future inner trapping horizon is a hypersurface foliated by marginally trapped surfaces, on which $\theta^{(\bm{l})}=0$ and $\mathscr{L}_{\bm{k}}\theta^{(\bm{l})}>0$. We just need to show that the second property is satisfied. However, this is just a corollary of the previous proposition and the structure of zeros of $\theta^{(\bm{l})}=0$ in a slice of constant $v$ that is necessary to avoid curvature singularities which, as discussed in Appendix \ref{sec:regconds}, must have $\mathscr{L}_{\bm{k}}\theta^{(\bm{l})}<0$ for the zeros in the $2k+1$ position where $m\geq0$ (with first one corresponding to the future outer trapping horizon), and $\mathscr{L}_{\bm{k}}\theta^{(\bm{l})}>0$ for the zeros in the $2k+2$ position. Regarding the spacelike character of the hypersurface $r=R_0(v)$, this just follows from the observation that this hypersurface lies outside the local light cone defined by outgoing and ingoing radial null geodesics. Indeed, let us consider a specific outgoing radial null geodesic reaching the point $\lambda=\lambda_0$ in which its expansion vanishes, and the ingoing radial null geodesic that crosses the same point towards smaller values of the radius, and therefore lies inside the mentioned outgoing radial null geodesic after the crossing point. As outgoing radial null geodesics cannot cross, $r=R_0(v)$ must lie outside the same outgoing radial null geodesic, and therefore outside the local light cone. $\blacksquare$

\vspace{0.2cm}

Let us now put all these pieces together in order to understand the global structure that is associated with the local geometric elements discussed in this section. In the following, $\mathscr{T}^3$ will denote the set of marginally trapped spheres (so that $\theta^{(\bm{l})}=0$):

\vspace{0.2cm}

\noindent
\textbf{Proposition 6:} $\mathscr{T}^3$ is a smooth compact submanifold only if the function $v(R_0)$ is not bijective, so that its domain contains at least two open subsets where the inverse $R_0(v)$ exists.

\noindent
\textbf{Proof:} Let us assume that $v(R_0)$ is bijective in all its domain of definition, so that its inverse $R_0(v)$ is strictly increasing and defines the (evolving) position of a future inner trapping horizon, as a consequence of propositions 4 and 5. In this situation, outgoing radial null geodesics stemming from trapped surfaces must intersect the trajectory of the future inner trapping horizon $r=R_0(v)$. These outgoing radial null geodesics, being inside the trapped region, have crossed before the future outer trapping horizon. Hence, all such geodesics must intersect $\mathscr{T}^3$ twice. If $\mathscr{T}^3$ is compact, there will always exist a ``tangent'' outgoing radial null geodesic intersecting $\mathscr{T}^3$ only once (this is the limiting case in between the geodesics with two crossings and the geodesics with no crossings). The alternative is that $\mathscr{T}^3$ is not compact, so that there is no outgoing radial null geodesic that intersects it only once (alternatively, one can take the limit in which this single intersection point is reached in infinite affine distance). In the compact case, both trapping horizons (inner and outer) will merge at the point where the ``tangent'' outgoing radial geodesic meets $\mathscr{T}^3$. Given that the position of the future inner trapping horizon is a strictly increasing function of the radial coordinate, $\mathscr{T}^3$ cannot be smooth at this point, displaying a cusp instead. $\blacksquare$

\vspace{0.2cm}

\begin{figure}[!h]
\begin{center}
\vbox{\includegraphics[width=0.25\textwidth]{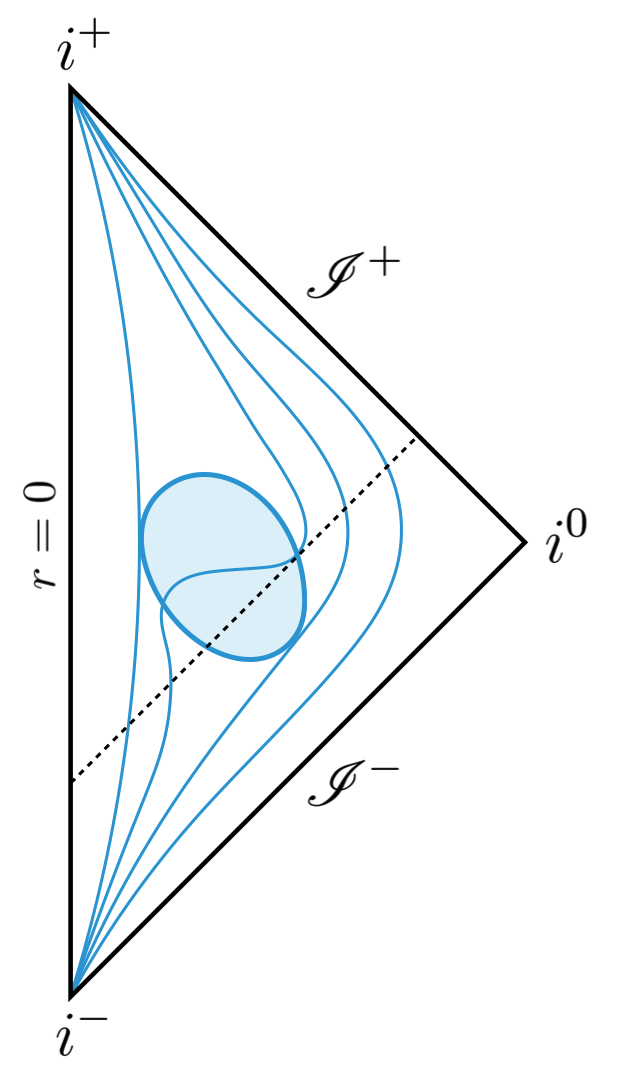}}
\bigskip%
\caption{Penrose diagram of the formation and dissapearance of a geodesically-complete black hole. The submanifold $\mathscr{T}^3$ is the boundary of the shaded region. The dashed line is an outgoing radial null geodesic, while the remaining curved lines mark the hypersurfaces of constant radius $r$.}
\label{fig:fig4}%
\end{center}
\end{figure}

\begin{figure}[!h]%
\begin{center}
\vbox{\includegraphics[width=0.55\textwidth]{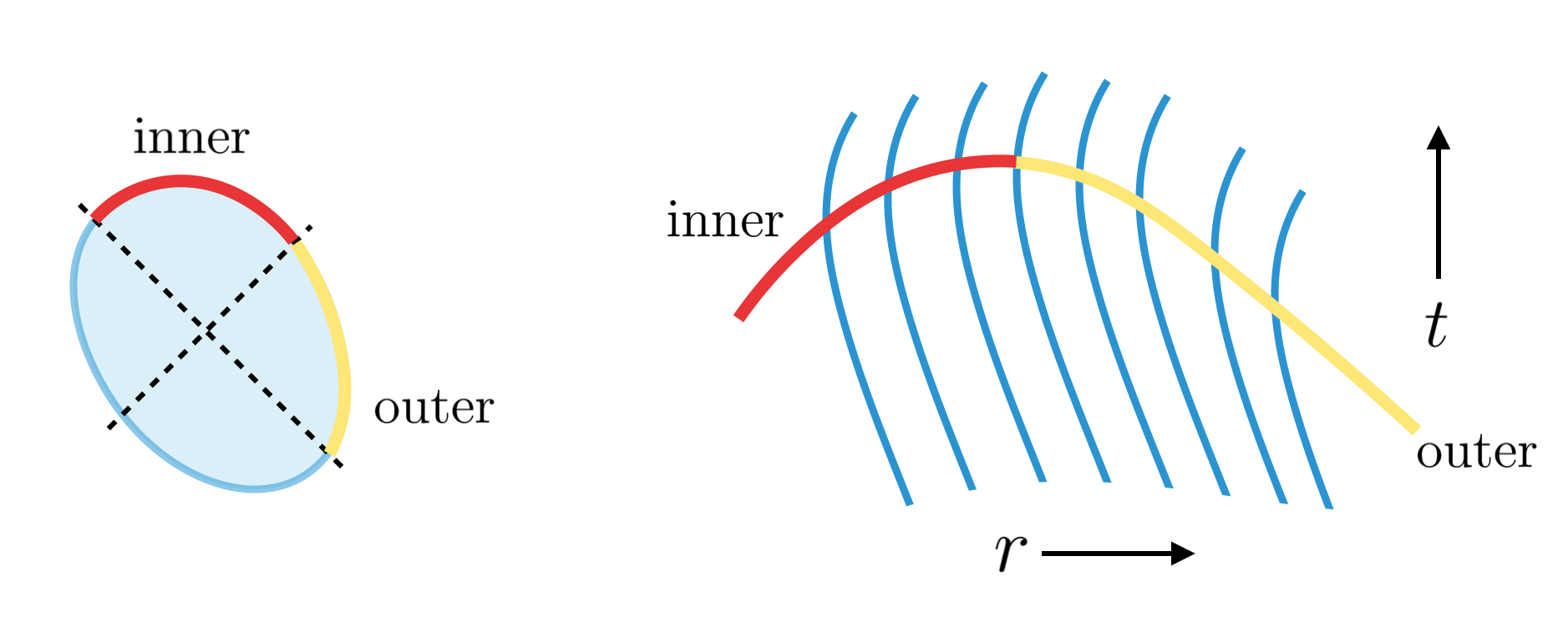}}
\bigskip%
\caption{Structure of the inner and outer future trapping horizons in the simplest spacetime that contains the ingredients in this section. To the best of our knowledge, the first explicit discussion of this structure was provided in \cite{Roman1983}. \emph{On the left:} the boundary of the top quadrant (red line) corresponds to the strictly increasing branch of $R_0(v)$, while the boundary of the right quadrant (yellow line) corresponds to the strictly decreasing branch of $R_0(v)$. \emph{On the right:} another perspective on the evolution of horizons, in which the hypersurfaces of constant $r$ are vertical lines and time flows in the upward direction.}
\label{fig:fig12}%
\end{center}
\end{figure}

The simplest situation compatible with the proposition above leads to the Penrose diagram depicted in Fig. \ref{fig:fig4} (more complicated situations just contain more outer and inner horizons). The topology of $\mathscr{T}^3$ is $S^1\times S^2$, and the domain of the function $v(R_0)$ has two subsets where an inverse exists. In one of these subsets (associated with the top quadrant on the left diagram in Fig. \ref{fig:fig12}), the function $R_0(v)$ is strictly increasing, so that this portion of $\mathscr{T}^3$ is identified with the expanding future inner trapping horizon. On the other hand, the right quadrant on the left diagram in Fig. \ref{fig:fig12} indicates the region in $\mathscr{T}^3$ in which $R_0(v)$ is decreasing, which is then identified as the shrinking future outer trapping horizon. Thus, the Penrose diagram depicted in Fig. \ref{fig:fig4} describes the formation and disappearance of a black hole that remains non-singular throughout its dynamical evolution, as forced by our assumptions. 

Our discussion does not assume any specific dynamical laws for either gravity or matter sectors which, in particular, implies that the disappearance of the black hole does not need to be linked with the emission of Hawking radiation. In other words, black hole evaporation would correspond to selecting specific geometries, with certain features and time scales, within the class that we have obtained as the result of our purely geometric analysis (moreover, it is important to keep in mind that we are not tracing explicitly the evolution of matter). For instance, an alternative to black hole evaporation that shares the same Penrose diagram was given in \cite{Rovelli2014}. It is also possible that $\mathscr{T}^3$ is not simply connected, but is composed of disconnected regions. Also, if one demands time-reversal symmetry, there must be a region of anti-trapped surfaces, and the corresponding spacetime describes the transition between a black hole and a white hole \cite{Barcelo2014a,Barcelo2014b,Haggard2014} (see Fig. \ref{fig:antitrapped}). 

While our geometrical analysis cannot tell us anything about the time scale of the process, the issue of mass inflation discussed in Sec.~\ref{sec:phen} implies that time-reversal symmetry is only compatible with the shortest possible time scale (roughly proportional to the light-crossing time); this is compatible with other classical and semiclassical instabilities \cite{Barcelo2015,DeLorenzo2015}. 
It is worth remarking that there is also a robust consequence of our assumptions for the dynamics of matter fields: the null energy condition must be violated. In fact, it has been proved in \cite{Hayward1993} that, if the null energy condition is satisfied by matter fields, the area of future outer trapping horizons must always increase. Hence, our general geometric assumptions are nevertheless tight enough to determine that the null energy condition (and, as a consequence, the dominant and weak energy conditions) must be violated deep down the gravitational well.

\begin{figure}[h]%
\begin{center}
\vbox{\includegraphics[width=0.45\textwidth]{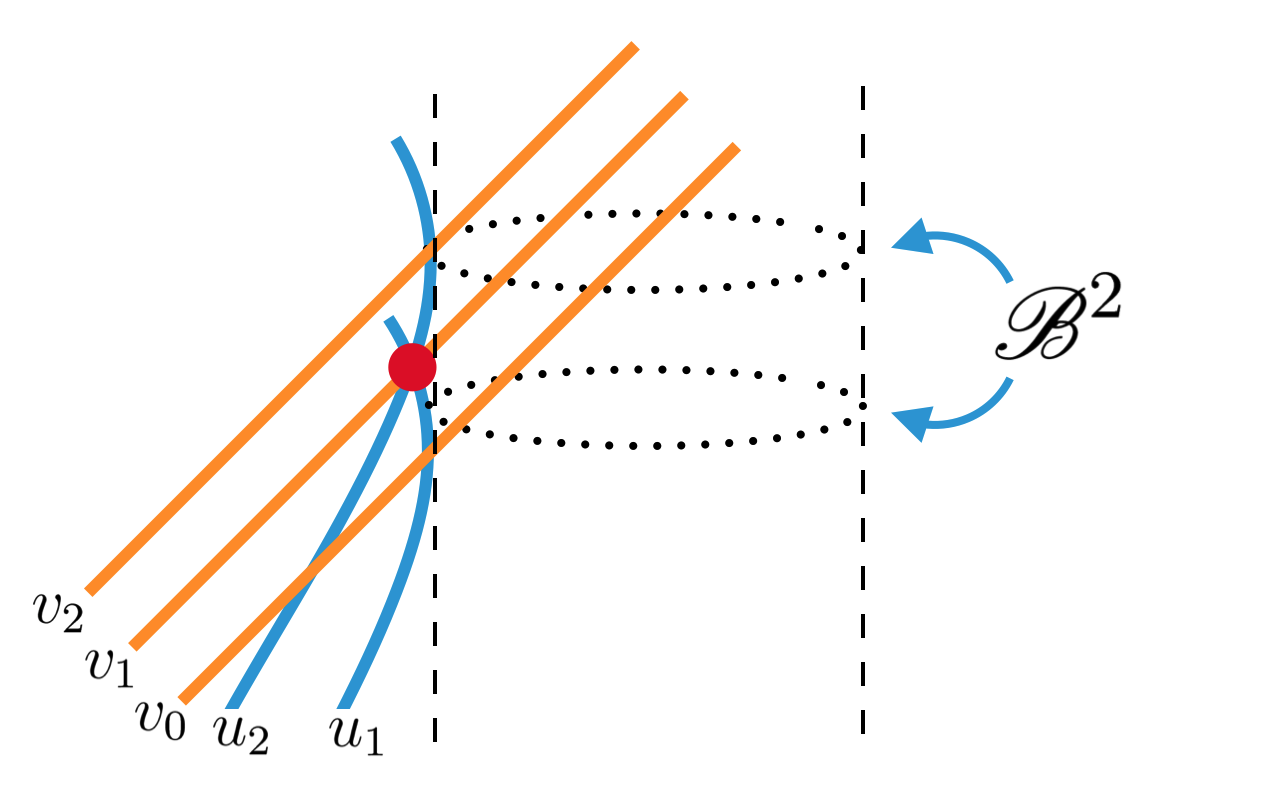}}
\bigskip%
\caption{Different coordinate patches that are relevant for our discussion on the avoidance of crossing points between outgoing radial null geodesics (the red dot in this figure). The straight lines at 45 degrees represent ingoing radial null geodesics.}
\label{fig:fig4b}%
\end{center}
\end{figure}%

Let us now show that the explicit assumption in the above discussion that there cannot be crossing points between two outgoing radial null geodesics is, in fact, a consequence of global hyperbolicity, starting with a detailed discussion of the case in which $R_0(v)$ is a constant function:

\vspace{0.2cm}

\noindent
\textbf{Proposition 7:} A spacetime in which $R_0(v)$ is a constant function cannot be globally hyperbolic.

\noindent
\textbf{Proof:} That there cannot be crossing points between two outgoing radial null geodesics having $\left.\theta^{(\bm{l})}\right|_{r=R_0}=0$ does not imply that these geodesics cannot have the same value of the radial coordinate for the corresponding values of their affine parameters. The (weaker) consequence that follows is that the one-to-one correspondence between the value of the radial coordinate and spheres in spacetime must break down, such that a given value of the radial coordinate does not uniquely identify a single sphere. We can try, of course, to use the set of null coordinates $(u,v)$ (in order to construct a complete set of coordinates we have to consider the angular coordinates as well, but it is not necessary to consider them explicitly). However, we shall now show that the geodesics reaching $r=R_0$ exit the spacetime region that can be covered with the $(u,v)$ coordinate patch. Indeed, let us assume that both spheres (located at the red dot in Fig. \ref{fig:fig4b}) can be covered with the $(u,v)$ patch. Then, one sphere will be labeled by the specific values $(u_1,v_1)$, while the other sphere will be associated with $(u_2,v_1)$. This would imply that the $u=u_1$ and $u=u_2$ geodesics must intersect $v=v_1$ at points that are identified with different values of the affine parameter along the latter ingoing radial null geodesic. However, this is in contradiction with Fig. \eqref{fig:fig4b}, as one can consider an open subset of the manifold around the red dot and follow the evolution of the affine parameter along the ingoing radial null geodesic in this subset (so that it must be the same for the two points labelled with different values of $u$). Given that null coordinates always exist locally, the second sphere must be labelled by a radial ingoing null geodesic that cannot cross the Cauchy surface $\mathscr{C}^3$. Hence, the spacetime is not globally hyperbolic. $\blacksquare$

\vspace{0.2cm}

Even if the proposition above is valid for $R_0(v)$ being a constant function, it is straightforward to see that a similar comment regarding the lack of global hyperbolicity in the presence of would-be intersection points applies to generic functions $R_0(v)$, as the argument above does not depend on the form of $R_0(v)$, but only on the interplay between different coordinate patches at the would-be interesection points. It is also worth stressing that the local behavior around these points may seem odd, but it is in fact ubiquitous in attempts of constructing non-singular black holes (we will discuss this in more detail in the corresponding cases below). The only difference is that our discussion is general enough to accommodate generic dynamical situations, while the well-known cases in the literature with Cauchy horizons are typically eternal black holes.\\
While we have so far focused on the minimal setting that just requires the existence of a single trapped region, as depicted in figure \ref{fig:fig4}, note that there are models proposed in the literature that present additional features while still belonging in this class. For instance, see Fig. \ref{fig:antitrapped} for a sketch of a situation in which there is a bounce and subsequent formation of an additional anti-trapped region \cite{Barcelo2014a,Barcelo2014b,Haggard2014}.
\begin{figure}
\begin{center}
\includegraphics[scale=.6]{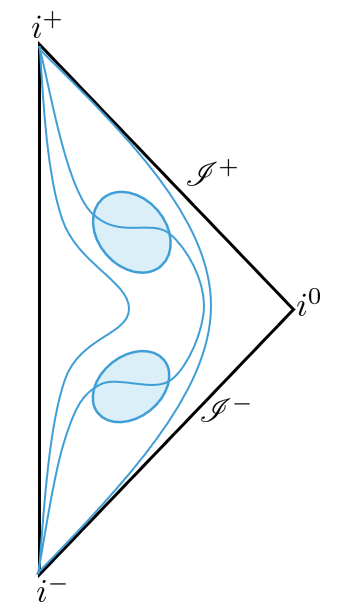}
\caption{After the disappearance of the trapped region, an anti-trapped region is formed. This situation describes a bounce from a black hole like geometry to a white hole like geometry. Situations in which a recollapse takes place \cite{Barcelo2014a,Barcelo2014b} can be described by a sequence of trapped and anti-trapped regions.}
\label{fig:antitrapped}
\end{center}
\end{figure}

\subsection*{Case A.II: One-way hidden wormholes $(\lambda_0,R_0,\bar{\theta}\geq0)$}\label{sec:wormholes}

Characterizing this class is simpler than for the previous one. We just need to take into account that there exists an ingoing radial null geodesic that starts with a negative value of the expansion, that eventually vanishes and becomes positive. That both expansions (outgoing and ingoing) vanish, although in different regions in spacetime, is characteristic of dynamical wormhole throats \cite{Hochberg1997,Hochberg1998}. In fact, we can show the following proposition:

\vspace{0.2cm}

\noindent
\textbf{Proposition 8:} There exists a hypersurface $r(u,v)=R_0$ where the radial coordinate has a local minimum.

\begin{figure}[!h]%
\begin{center}
\vbox{\includegraphics[width=0.8\textwidth]{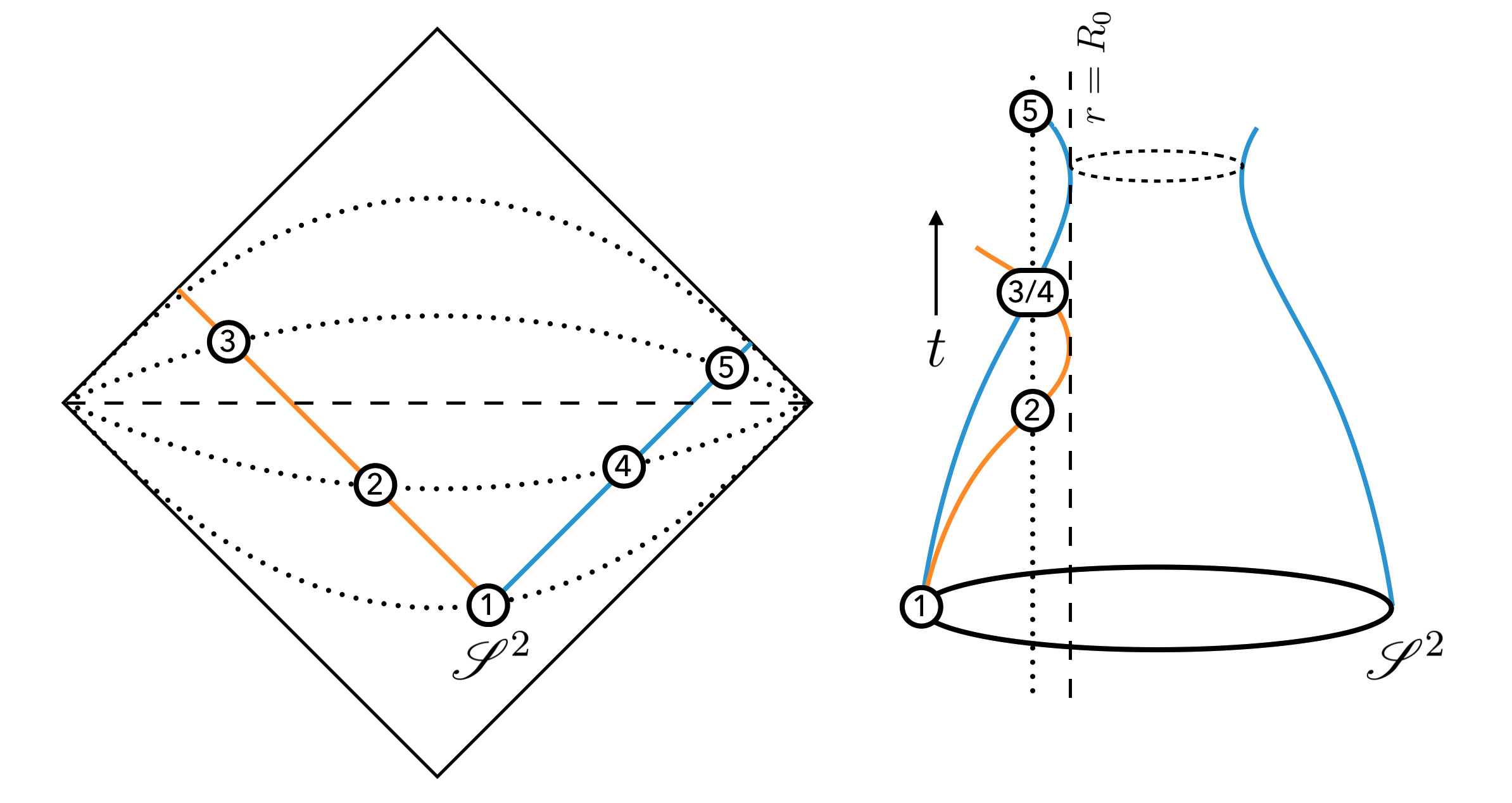}}
\bigskip%
\caption{Part of the Penrose diagram of a static wormhole inside a future outer trapping horizon, corresponding to the quadrant containing the throat of the wormhole. The straight dashed line indicates the line of constant $r$ with the minimum value of the latter, while the dotted lines are also lines of constant $r$ but associated with with greater values. Two radial null geodesics, ingoing and outgoing, depart from the crossing point 1. The ingoing geodesic crosses the point 2 with a value of the radius corresponding to the would-be crossing point 3, reaches $r=R_0$ (which is the same for both ingoing and outgoing geodesics due to the wormhole being static), and bounces back until reaching what looks like a second crossing point 3. However, as illustrated by the Penrose diagram, there is no actual crossing, as the outgoing geodesic would be when having this same radius at point 4 of spacetime. That is, the right-hand figure, being based on using the $r$ coordinate as primary, is misleading -- exactly because the $r$ coordinate is double-valued in this class of spacetimes.}
\label{fig:fig6}%
\end{center}
\end{figure}%

\noindent
\textbf{Proof:} The value of the radial coordinate along the outgoing radial null geodesic stemming from $\mathscr{S}^2$ (see the right panel in Fig. \ref{fig:fig6}) must reach a minimum at $r=R_0$. Let us now study the intersections between outgoing and ingoing radial null geodesics, in similar terms as we did before the intersection of two outgoing radial null geodesics. Instead of dealing with a single intersection between outgoing radial null geodesics, in this case we deal with double intersections between outgoing and ingoing radial null geodesics (Fig. \ref{fig:fig6}). This is also forbidden in spherical symmetry, as it would imply that two different points would be associated with the same values of (null) coordinates. In fact, as in the previous case, both coordinate sets $(u,r)$ and $(v,r)$ cannot be used to go through the points in which the expansion of ongoing and ingoing null geodesics, respectively, vanishes. However, the main difference with respect to the previous case is that the splitting of the second intersection point can be well described in the $(u,v)$ coordinates, as illustrated graphically in Fig. \ref{fig:fig6}, but at the price of creating two different hypersurfaces in spacetime corresponding to the same value of the radius (along the points 2, 4 and 3, 5 in the figure, respectively). Hence, when other trapped surfaces (and the corresponding outgoing radial null geodesics) in the neighborhood of $\mathscr{S}^2$ are considered, it follows that there must be a hypersurface $r(u,v)=R_0$ in which the radial coordinate has a local minimum. $\blacksquare$

\vspace{0.2cm}

\begin{figure}[h]%
\begin{center}
\vbox{\includegraphics[width=0.55\textwidth]{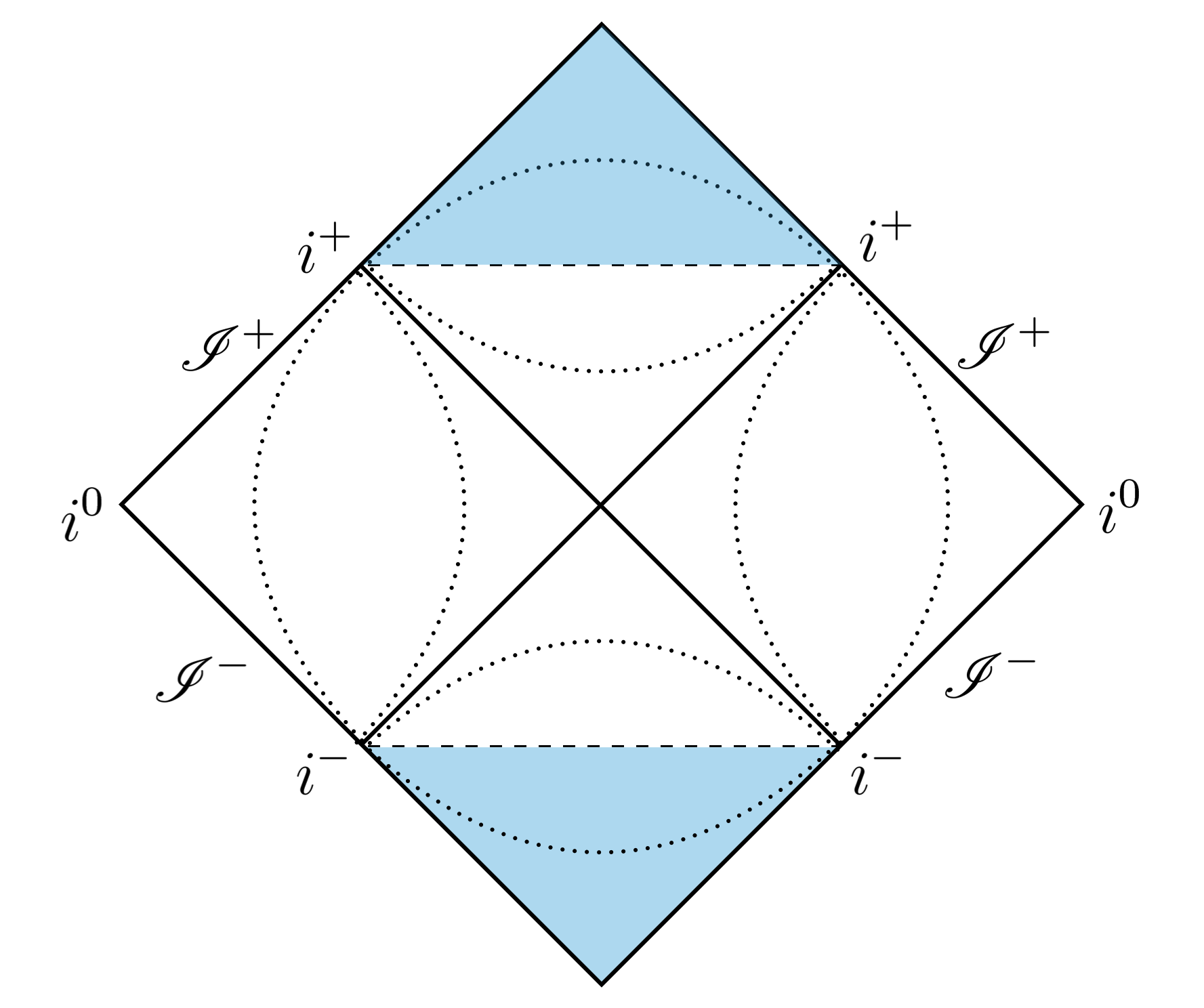}}
\bigskip%
\caption{Dashed lines are lines of minimum radius $r=R_0$ (corresponding to $\mathscr{B}^2$ in Fig. \ref{fig:fig2}). Dotted lines are lines of constant radius. The shaded regions indicate points in the manifold that cannot be covered with either the $(u,r)$ or $(v,r)$ coordinate systems.}
\label{fig:fig7}%
\end{center}
\end{figure}%

This situation generally corresponds to a one-way (hidden) dynamical wormhole \cite{Hochberg1998}, except for the limiting case $\left.\theta^{(\bm{k})}\right|_{\lambda=\lambda_0}=0$ which describes a static wormhole \cite{Hochberg1997}. It is important to keep in mind that these are not wormholes in the most common sense of this term, as the throats are inside a future outer trapping horizon. 

\subsection*{Case B.I: Everlasting horizons $(\infty,R_\infty,\bar{\theta}<0)$}

\begin{figure}[!h]%
\begin{center}
\vbox{\includegraphics[width=0.25\textwidth]{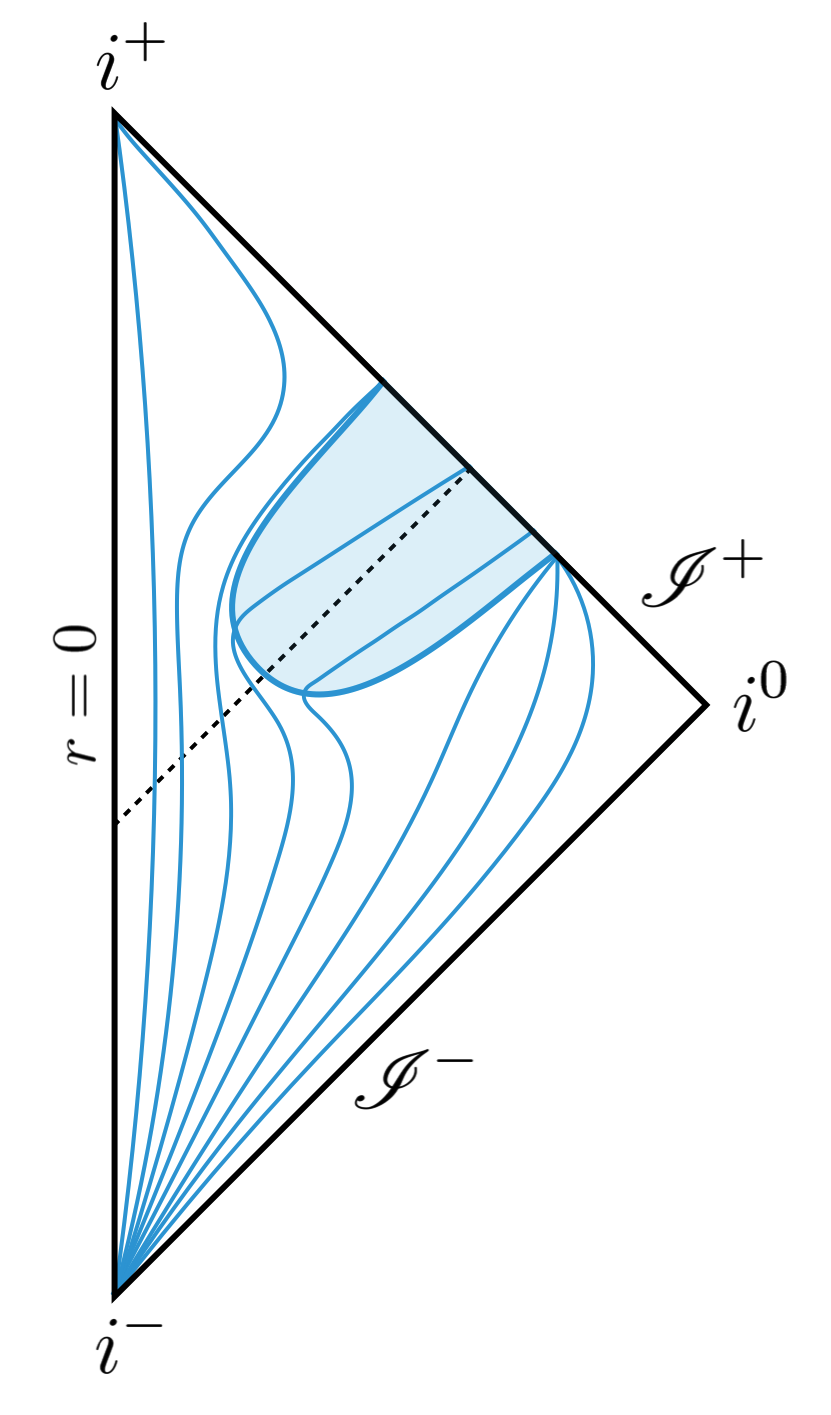}}
\bigskip%
\caption{A geodesically-complete black hole that forms during gravitational collapse but never disappears, and without the formation of a Cauchy horizon.}
\label{fig:fig10}%
\end{center}
\end{figure}%

This situation can be obtained as the $\lambda_0\rightarrow\infty$ limit of $(\lambda_0,R_0,\bar{\theta}<0)$. The toroidal topology of the horizons in the latter case is deformed so that outgoing radial null geodesics never reach the second point (along their trajectories) in which $\theta^{(\bm{l})}$ vanishes. The corresponding Penrose diagram, provided in Fig. \ref{fig:fig10}, is a variation of the Penrose diagram in Fig. \ref{fig:fig4}. In physical terms, the diagram in Fig. \ref{fig:fig10} is an intermediate situation between a static regular black hole and Fig. \ref{fig:fig4}, in which outgoing radial null geodesics do not reach either $r=0$ or $r=\infty$ (after crossing the future inner trapping horizon), but rather an intermediate radius $r=R_\infty(u)$. The corresponding black hole is always growing and never disappears; any process that describes the partial evaporation or total disappearance of the black hole must involve some outgoing radial null geodesics satisfying the conditions characteristic of the case $(\lambda_0,R_0,\bar{\theta}<0)$ studied above.

Another issue that is worth clarifying is the asymptotic ($\lambda\rightarrow\infty$) behavior of the lines of constant radial coordinate in the trapped region which, as depicted in Fig. \ref{fig:fig10}, become null asymptotically. This can be seen to be a consequence of $\theta^{(\bm{l})}$ vanishing in the limit $\lambda\rightarrow\infty$, which implies that the function $F(v,r)$ in Eq. \eqref{eq:linel} vanishes as well (see App. \ref{sec:regconds} for explicit expressions), and so does the line element on the lines of constant radial coordinate,
\begin{equation}
\left.\text{d}s^2\right|_{r=r_0}=-e^{-2\phi(v,r)}F(v,r)\text{d}v^2.
\end{equation}
%

\subsection*{Case B.II: Asymptotic hidden wormholes $(\infty,R_\infty,\bar{\theta}\geq0)$}

Similarly as the case just above, this situation can be understood as the $\lambda_0\rightarrow\infty$ limit of a previously discussed case, namely $(\lambda_0,R_0,\bar{\theta}\geq0)$, which described a wormhole. Hence, this kind of geometry corresponds to an asymptotically cylinder-like geometry. This is easiest to realize for the static case $\lim_{\lambda\rightarrow\infty}\theta^{(\bm{k})}=0$, in which the radius of the cylinder is given by the size of the corresponding surface $\mathscr{B}^2$ (which is at infinite affine distance). The Penrose diagram would be given by Fig. \ref{fig:fig7} but removing the shaded regions. In the dynamical situation, we would have a ``half-wormhole'' where the second universe has been discarded and the throat pushed out to infinite affine distance along outgoing radial null geodesics.

\begin{figure}
\begin{center}
\includegraphics[scale=0.8]{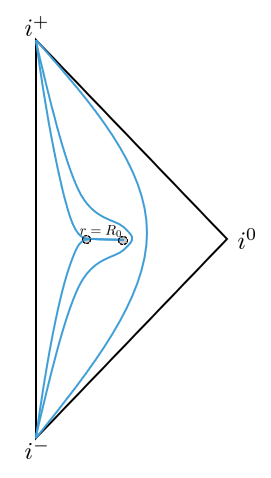}
\caption{One-way hidden wormhole with some lines of constant $r$ depicted. In order to avoid issues with topology change the metric must fail to be analytic in at least two points. Such points are encircled for clarity.}
\label{fig:nonanalytic}
\end{center}
\end{figure}

Before addressing the classification of the remaining  classes of geometries let us pause to note that it is very difficult to make the one-way hidden wormhole geometries described in this section and by the class $(\lambda_0,R_0,\bar{\theta}\geq 0)$ consistent our hypotheses. In fact, if we only consider geometries described in term of analytical metrics, this is only compatible  with the topology being $\mathbb{R}^3$ in the past if we assume the existence of a spacelike boundary for the manifold.  The fact that the boundary has to be spacelike is simple to understand as global hyperbolicity implies that all the Cauchy hypersurfaces reach $r=0$. On the other hand, the presence of the boundary is necessary as otherwise the $r=R_0$ line, besides having a portion which is a minimum radius hypersurface (that is, the radius would be greater at both sides of the hypersurface), would have to reach both $i^-$ and $i^+$. If we trace this $r=R_0$ line back to $i^-$, one goes continuously from a situation in which the radius is greater at both sides of this hypersurface to a situation in which the radius is greater on one side and smaller on the other one. Let us focus our attention on the side in which there is a transition from the radius being greater than its value on this hypersurface to becoming smaller, and take two points in these two regions. Along an arbitrary line connecting these two points, it is clear that, due to continuity, there is a point in which $r=R_0$ as well (which is not contained in the previous hypersurface). This argument can be repeated for all points on this side of the original $r=R_0$ hypersurface and arbitrarily close to the latter, which implies the existence of two $r=R_0$ that merge at some point. The metric cannot be analytical at this point. Also, the single $r=R_0$ line that remains after the merger must disappear (if we assume that it does not end up in $i^0$), which implies the existence of another point in which the metric is non analytic (see Fig. \ref{fig:nonanalytic}  for a sketch of the Penrose diagram in this situation). \\
Finally let us note that there is one last possibility that have been considered to make this class of geometries viable. In \cite{Bianchi:2018} the authors provide a metric which locally describes an  element of this class of geometries but which is not globally defined, \textit{i.e.}, there must exist a region where an effective spacetime is not defined (these regions are related to non-analyticity issue described before).  This possibility lies outside our classification as we are assuming an everywhere defined metric. 

%

\subsection*{Cases B.III, B.IV, C.I and C.II: Curvature singularities, $(\infty,0,\bar{\theta}<0)$, $(\emptyset,0,\bar{\theta}<0)$, $(\infty,0,\bar{\theta}\geq0)$ and $(\emptyset,0,\bar{\theta}\geq0)$}

These four situations are sufficiently similar that we can analyze them simultaneously, with only slight differences due to the different behavior of the expansion of ingoing radial null geodesics. In fact, their main feature is that outgoing radial null geodesics reach $r=0$. It is straightforward to show that this is not compatible with the regularity of curvature invariants, provided we maintain all our assumptions (such as geodesic completeness). Outgoing radial null geodesics reach $r=0$ in an infinite affine distance but, generally, ingoing null geodesics will reach $r=0$ (as there is no wormhole throat in this case) in finite affine distance. At the same time, there is no future inner trapping horizon (in other words, a second point in which $\theta^{(\bm{l})}$ vanishes) which, following the discussion in Sec. \ref{sec:regconds}, is a necessary condition to ensure regularity at $r=0$. Hence, these cases lead to a curvature singularity at infinite affine distance along outgoing null geodesics, but generally finite affine distance along ingoing null geodesics for instance. In other words, we have seen in the discussion of the cases above that, in order to avoid a curvature singularity at $r=0$, one needs either a wormhole structure (a minimum value of the radial coordinate that prevents reaching $r=0$) or a future inner trapping horizon. As the geometries studied in this item contain by construction none of these, these must be singular (if global hyperbolicity is maintained). 

\begin{figure}[!h]%
\begin{center}
\vbox{\includegraphics[width=0.45\textwidth]{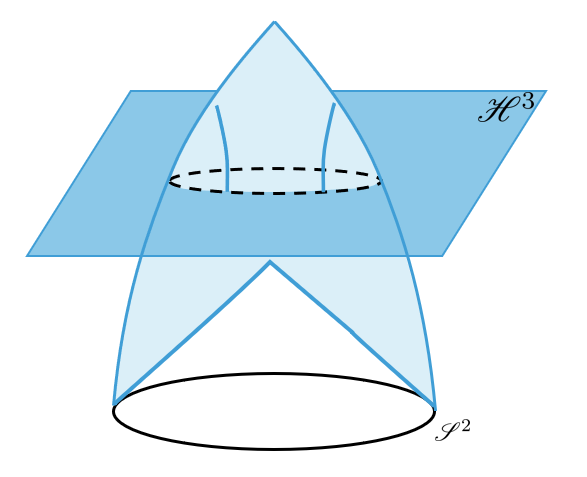}}
\bigskip%
\caption{Static regular black holes evade Penrose's theorem due to the introduction of a Cauchy horizon (here represented as the horizontal plane) before the focusing point is reached, which allows the focusing point to be regular in these geometries.}
\label{fig:fig8}%
\end{center}
\end{figure}%

\begin{figure}[!h]%
\begin{center}
\vbox{\includegraphics[width=0.5\textwidth]{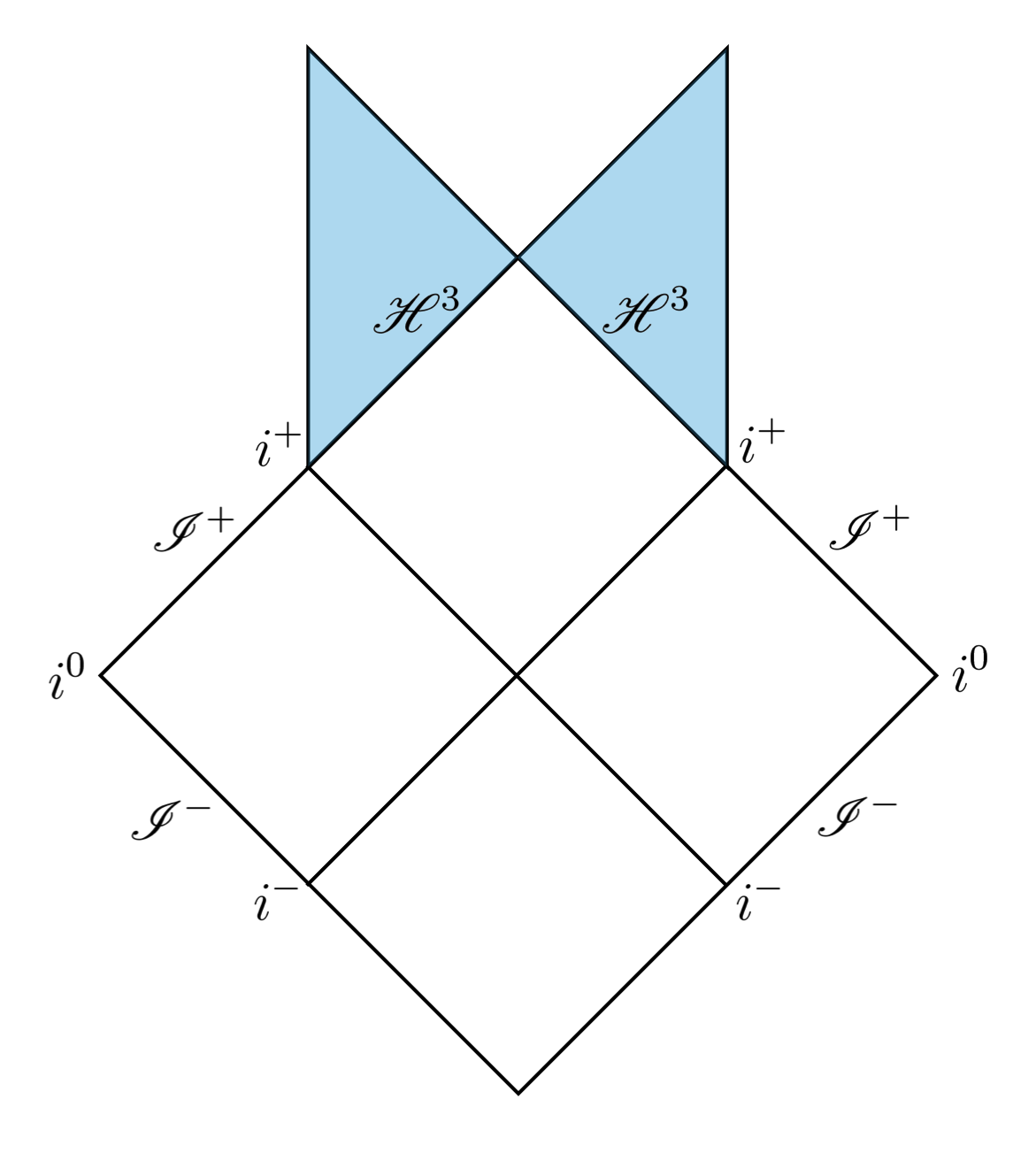}}
\bigskip%
\caption{Penrose diagram that can represent different kinds of (eternal) black holes, from charged to rotating black holes, but also regular black holes in which the central singularity has been removed. The colored quadrants on the left and on the right represent respectively the regions which are covered by the $(v,r)$ patch but not with the $(u,r)$ patch, and by the $(u,r)$ patch but not with the $(v,r)$ patch. These are particular examples of spacetimes that are not globally hyperbolic, following our general discussion in this section. The complete Penrose diagram is obtained by stacking indefinitely the section represented here. Our discussion in this paper is more general, as it does not require that the diagram is not symmetric as this one (this symmetry is a consequence of the eternal nature of the black holes represented in this diagram).}
\label{fig:fig5}%
\end{center}
\end{figure}

It is always possible to violate the assumption of global hyperbolicity in Penrose's theorem (which is also one of the four conditions behind our analysis) in order to avoid being singular. Interestingly, geometries that appear often in the literature belong to this case. Even if the focusing point in these situations is reached generically in finite affine distance, this point in the discussion is a natural placement of these well-known geometries describing static regular black holes \cite{Bardeen1968,Frolov1979,Frolov1981,Roman1983,Dymnikova1992,Borde1996,Hayward2006,Bronnikov2006,Dymnikova2001,Ansoldi2008,Frolov2014,Frolov2016,Carballo-Rubio2018}. As shown graphically in Fig. \ref{fig:fig8} , these geometries do not evade Penrose's theorem by avoiding the formation of focusing points, but rather by introducing a Cauchy horizon before this focusing point is reached by outgoing geodesics. For completeness, we also provide in Fig. \ref{fig:fig5} the Penrose diagram of the corresponding eternal geometries. Static regular black holes correspond to the limiting situation in which the expansion $\theta^{(\bm{k})}$ associated with the ingoing radial null geodesics crossing $\mathscr{C}^3$ vanishes along the Cauchy horizon (in these spacetimes, there are ingoing radial null geodesics that intersect the trapped outgoing radial null geodesics while having $\theta^{(\bm{k})}$, but only in the region beyond the Cauchy horizon). One may think that it should be possible to devise situations in which $\theta^{(\bm{k})}$ actually changes sign without violating global hyperbolicity. However, these situations would always involve two intersections between outgoing and ingoing radial null geodesics which, due to the lack of wormhole throats in this situation ($r=0$ is reached by outgoing radial null geodesics by construction), implies the lack of global hyperbolicity. 

\section{Self-consistency and phenomenology}\label{sec:phen}

As a follow-up to Penrose's theorem, we have discussed the possible families of spacetimes that, while containing closed trapped surfaces, remain geodesically complete and free of curvature singularities. Our analysis is geometric in nature and therefore does not address the dynamical laws that are needed in order to construct specific realizations of these families. However, we can conclude that any theory beyond general relativity leading to spacetimes satisfying conditions 1 to 4 in Sec. \ref{sec:penrth} must yield a member of one of these families (or, at most, a hybrid combination).

Hence, this result serves as an intermediate layer that may help bridging the gap between theories of quantum gravity and observational physics. These families of spacetimes can be understood as effective descriptions of the evolution of non-singular black holes (defined as regions containing closed trapped surfaces but without singularities), sharing some qualitative properties but still displaying quantitative differences that can be described in terms of parametric and functional degrees of freedom. From this intermediate layer, one can work towards embedding at least some of these families in a given theory of quantum gravity, thus reducing the aforementioned freedom from a theoretical perspective, or turn to analyzing the phenomenological implications that follow in order to constrain these degrees of freedom observationally (and, perhaps, even discard some of these families).

In this final discussion, we want to highlight that the effective description in terms of these spacetimes would only be meaningful if displaying a satisfactory level of self-consistency, and discuss the phenomenological implications that follow from this requirement. Of course, the lack of knowledge of the specific dynamics that may yield these spacetimes implies a degree of uncertainty about some fine details of these models. However, this does not prevent extracting certain conclusions that, if formulated carefully enough, would remain robust regardless of this uncertainty.

One generic property of many of the families of geometries discussed in this paper is the presence of instabilities. These instabilities are associated with the well-known phenomenon of mass inflation in general relativity \cite{Simpson1973,Poisson1989,Poisson1989b,Poisson1990,Barrabes1990,Ori1991,Markovic1994,Dafermos2003,Hamilton2008,Cardoso2017,Hod2018,Dias2018}, and have been recently discussed for generic geometries describing a particular kind of non-singular black hole \cite{Carballo-Rubio2018} corresponding to one of our families in this paper, namely $(\lambda_0,R_0,\bar{\theta}<0)$ with the Penrose diagram depicted in Fig. \ref{fig:fig4}. In more strict terms, the discussion in \cite{Carballo-Rubio2018} was focused on static situations in which the Penrose diagram would look like that in Fig. \ref{fig:fig5}. However, it is not difficult to show that the main ingredients arising in the discussion of the static cases carry over the dynamical situation (thus indicating that the discussion of static situations is indeed representative of more general situations, thus addressing the warnings raised in \cite{Schindler2019}). There is no need to repeat here the calculations that were performed in the latter reference, but we would like to highlight the ingredients that allow us to proceed in a similar fashion. This can be done easily on the basis of our explicit discussion of the behavior of outgoing and ingoing radial null geodesics.

\begin{figure}[!h]%
\begin{center}
\vbox{\includegraphics[width=0.35\textwidth]{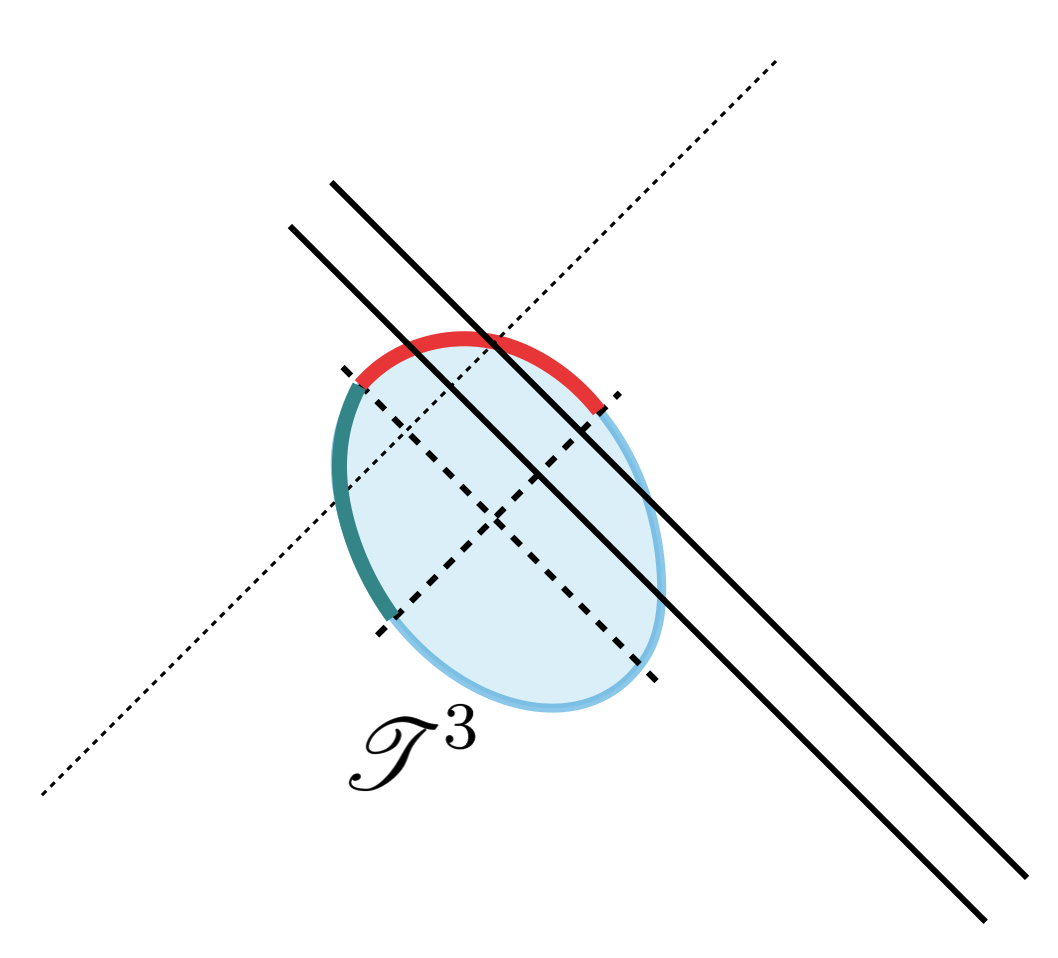}}
\bigskip%
\caption{Detail of the trapped region in Fig. \ref{fig:fig4} and an outgoing radial null perturbation (dashed line) that intersects several ingoing radial null perturbations.}
\label{fig:massinf}%
\end{center}
\end{figure}%

Following the discussion in \cite{Carballo-Rubio2018}, let us consider a toy model in which the perturbations that will awaken the instability are null thin shells, following the outgoing and ingoing radial null geodesics that forms the basis for our classification. We can then consider the effect of these perturbations on diagrams such as Fig. \ref{fig:fig3}, considering for simplicity a single outgoing shell and several ingoing shells that intersect at some point in the future (see Fig. \ref{fig:massinf}). We know from our discussion in Sec. \ref{case:1} that outgoing radial null geodesics inside the trapped region will reach in finite affine distance a point in which their expansion vanishes. This point is given by the second intersection between the outgoing radial null geodesic depicted in Fig. \ref{fig:fig4} and the submanifold $\mathscr{T}^3$ (the first intersection just marks the entrance of this outgoing radial null geodesics in the trapped region). This allow us to highlight the main difference between the dynamical and static cases: in the static case the expansion $\theta^{(\bm{l})}$ never vanishes, but $\theta^{(\bm k)}$ does along the Cauchy horizon. However, this difference is inconsequential for our present discussion as, regardless of the specific expansion vanishing, in both cases the function $F(v,r)$ in Eq. \eqref{eq:linel} vanishes. Hence, whenever the crossing points between outgoing and ingoing radial null perturbations are close (from inside) to the upper quadrant of $\mathscr{T}^3$ (the inner part in Fig. \ref{fig:fig12}), we can then use a Taylor expansion of this function in order to calculate the mass in the different quadrants of spacetime defined by the two crossing shells, and integrate the resulting equations along the outgoing radial null geodesic describing the outgoing perturbation. It follows then that the mass in the quadrant that lies in the causal future of the crossing point grows exponentially with the value of $v$ associated with the ingoing shell, being proportional to
\begin{equation}\label{eq:expg}
e^{-|\bar{\kappa}_-|\Delta v},
\end{equation}
where $\bar{\kappa}_-$ is the average value on the interval $\Delta v$ of the surface gravity of the future inner trapping, defined for the metric in Eq. \eqref{eq:linel} as 
\begin{equation}
\kappa_-(v)=\left.\frac{e^{-\phi(v,r)}}{2}\;\frac{\partial F(v,r)}{\partial r}\right|_{r=R_0(v)}.
\end{equation}
The only condition that is necessary in order to obtain Eq. \eqref{eq:expg} is that $R_0(v)$ remains roughly constant during the interval $\Delta v$, so that its variation can be neglected. Hence, there is an instability if the typical time scales of variation of $R_0(v)$ is much larger than $1/|\bar{\kappa}_-|$. This reasoning applies to the portion of the inner horizon in the upper quadrant of Fig. \ref{fig:massinf}, for which the intersection point of one ingoing and  several outgoing null shell approaches the horizon in the future. A similar analysis can be performed for the other portion of the inner horizon in the left quadrant of Fig. \ref{fig:massinf}, considering one outgoing and several ingoing null shells obtaining a instability exponentially growing in $u$.

These equations reduce to the ones derived in \cite{Carballo-Rubio2018} in the static limit. A straightforward calculation shows that the exponential growth encapsulated in Eq. \eqref{eq:expg} is also reflected in curvature invariants. The typical time scale of this instability is $1/|\bar{\kappa}_-|$; unless this quantity is fine-tuned, its natural value is essentially given by the coordinate light-crossing time of the sphere $r=r_{\rm inner}(v)$ which, in scenarios motivated by quantum gravity, is roughly the Planck time \cite{Carballo-Rubio2018}. Hence, this time scale is typically extremely short, if compared with the standard time scale associated with Hawking evaporation (that is cubic in the mass of the black hole). Of course, once a full quantum gravity description is available, it might be the case that what is identified as fine-tuning is actually the only possibility that is realized. However, without this knowledge, and limiting our discussion to the effective description considered in this paper (and other works dealing with regular black holes), the only possible way to deal with this instability is through extreme fine-tuning\footnote{That is, it is necessary to tune the order of magnitude dimensionless numbers to ensure that black holes can be long-lived.} of the properties of spacetime.

For completeness, let us mention that the proof above is based on null perturbations so that, in strict terms, it cannot be used to infer the existence of instabilities in certain situations in which outgoing radial null geodesics do not find any inner horizon. This kind of situation can be constructed as a limiting case of the discussion in Sec. \ref{case:1} in which the domain of trapped surfaces $\mathscr{T}^3$ is not compact. It is always possible to take one point in the boundary of $\mathscr{T}^3$ and displace it to $\mathscr{I}^+$ in a way that outgoing radial null geodesics to not meet the inner horizon (the subsequent $\mathscr{T}^3$ has the form of a drop in the Penrose diagram). However, timelike perturbations will always cross the inner horizon, and it is possible to perform a similar argument using timelike instead of null shells. Hence, there is no reason to expect that this would ameliorate the issue of mass inflation. Moreover, this case describes a black hole that never disappears, which poses other self-consistency problems as described below.

A possible way out of the instability issue is assuming the existence of other dynamical processes that become more important than Hawking evaporation, causing the disappearance of trapping horizons in shorter time scales \cite{Barcelo2015}. Interestingly, the instability time scale being typically Planckian implies that time scales that are linear in the mass (as proposed in \cite{Barcelo2014a,Barcelo2014b}) would not be plagued by this self-consistency issue. However, the price one has to pay is a radical change on our perspective of the actual nature of astrophysical black holes. Whether this price is worth paying would possibly depend on personal preconceptions and, moreover, it is not an issue we want to discuss here; for the purposes of the present analysis, it is enough to stress that this possibility would most likely entail a number of distinctive phenomenological signatures \cite{Carballo-Rubio2018b}.

It is remarkable that the only family of geometries that describes situations in which black holes disappear in a finite amount of time, namely $(\lambda_0,R_0,\bar{\theta}<0)$, suffers from this instability. The remaining families describe situations with a final phase in the evolution of black holes characterized by the formation of an asymptotic state still containing trapping horizons: a remnant. These scenarios have been analyzed in the literature, especially in connection with the information loss problem, and the consensus seems to be that remnants face self-consistency issues that require further analysis \cite{Giddings1993a,Giddings1993b,Susskind1995,Chen2014}. (See also the corresponding discussion in \cite{Hossenfelder2009}.) Just for completeness, let us mention that these criticisms are based either on the observation that it does not seem possible for remnants to store the amount of information that would be necessary in order to solve issues with information loss, or with estimates of their production cross-section that yield extremely large values. 

Let us focus, however, on the conclusions that can be drawn from our purely geometric analysis, and the assumptions behind it. First of all, we should mention that some of the classes in this category do also present important conceptual issues regarding their formation in realistic collapse scenarios: for instance, $(\lambda_0,R_0,\bar{\theta}\geq0)$ requires topology change to take place, which may obstruct its viability as a complete description of the formation and evolution of black holes (see, however, \cite{Hsu2006}). On the other hand, $(\infty,R_0,\bar{\theta}<0)$ is also affected by the mass inflation instability, although for this family it is less clear whether fine-tuning is strictly necessary in order to avoid this issue.

Another ingredient that is not present into our geometrical analysis but that is expected to be satisfied by a realistic scenario is the presence of Hawking radiation in agreement with semiclassical physics at least in some regimes. The classes $(\infty,R_\infty,\bar{\theta}\geq0)$ and $(\infty,R_\infty,\bar{\theta}<0)$  can be compatible with the existence of Hawking radiation if and only if the latter eventually switches off due to some unknown quantum gravity effect. However, accepting that this is the case is tantamount to dropping our assumption that it is possible to construct a meaningful description of the complete dynamical evolution of black holes that satisfies our assumptions (1--4). Indeed, the switching off of Hawking radiation cannot be described using quantum field theory on the geometric backgrounds leading to remnants that have been discussed in complete generality in this paper; this description must break down at some point, meaning that the effective description in terms of differentiable manifolds is no longer meaningful below a certain length scale.

\section{Conclusions and discussion \label{sec:discussion}}

We have studied in completely general terms the possible spherically symmetric geometries describing geodesically-complete black holes. Beside geodesic completeness, we have also demanded global hyperbolicity (so that it is reasonable to think that these geometries can be obtained through dynamical evolution) and the absence of curvature singularities.

\begin{figure}[h]
\begin{center}
\includegraphics[scale=.65]{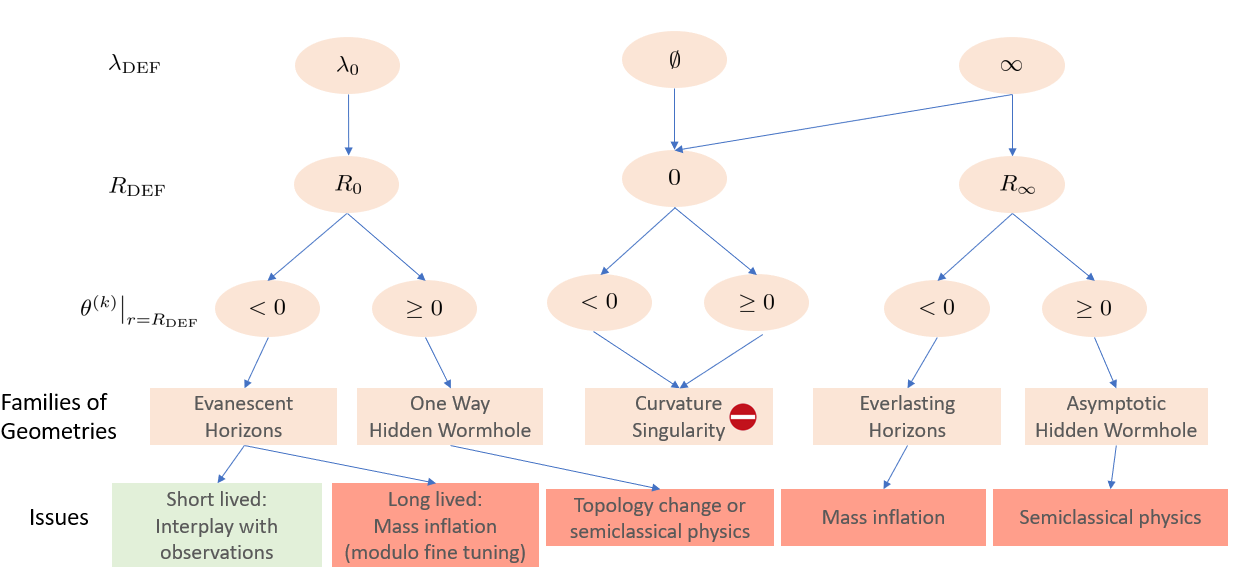}
\caption{The table summarizes the classes of the allowed geometries. The most general geometry is given by a combination of the geometries in the table. }
\label{fig:diagram}
\end{center}
\end{figure}

Remarkably, we have found that purely geometric considerations are enough to strongly constrain the allowed class of spacetimes. Using our classification, we were able to identify four classes of geometries, that can either be taken as they are or be combined, in order to yield the most general spherically symmetric spacetime satisfying our requirements (1--4). Figure \ref{fig:diagram} summarizes the classification that we have provided and the different viable classes of geometries.

The first class corresponds to geometries in which the defocusing of outgoing null geodesics is placed at a finite value of the affine parameter and for a finite value of the radial coordinate, whereas ingoing null geodesics do not have a defocusing point. This class of geometries is associated with a geodesically-complete black hole that disappears in finite time, which we call an ``evanescent regular black hole". Note that we are carefully avoiding the use of ``evaporation"; our geometric analysis does not provide any hint about the dynamical process behind the disappearance of the black hole. Such process could either be Hawking radiation or some other process taking place in shorter time scales, for instance the proposed transition to a white hole; as explained in Sec. \ref{sec:phen}, we can also conclude that long time scales are disfavored due to the mass inflation instability at the inner horizon.

The second class differs from the first one due to the value of the expansion parameter of ingoing null geodesics, which is non-negative at the point in which the defocusing point for the outgoing geodesics. We have discussed that this implies the existence of something closely resembling a wormhole throat. We refer to these geometries as ``one-way hidden wormholes" to stress the fact that the throat is inside a trapped region. One must keep in mind that this possibility requires a change of topology during stellar collapse. Even if it is not possible to exclude this possibility as we do not know yet the fundamental quantum gravity theory, it is also not possible to embrace this possibility without further questioning, as it is clear that this issue must be solved before considering this possibility a realistic one. We have also discussed that this situation clashes with the semiclassical prediction of Hawking radiation.

The last two classes can be understood as the limit of $\lambda_\defocus\rightarrow\infty$ of the previous two cases. The difference between these two classes is once again the value of the expansion parameter of ingoing null geodesics. If the expansion is negative, then we have an ``everlasting regular black hole". It is difficult to reconcile this picture with the mass inflation instability previously discussed. If the expansion parameter is asymptotically zero or positive, we would get an ``asymptotic hidden wormhole" which would again be non-traversable and now involve a one-way trip to the edge of spacetime. These classes are also not compatible with semiclassical physics, in the sense that evaporation due to Hawking radiation cannot be complete.

Overall, we find the remarkable result that there is no fully ``safe" option: all classes of geometries display internal inconsistencies (instabilities, topology change or incompatibilities with semiclassical physics), except for the particular case of evanescent black holes in short time scales. However, the time scales in the latter case are so short (of the order of the light-crossing time) that this possibility entails a complete rethinking of the concept of black hole. Indeed, a fast bounce would not be obviously at odds with observations only if the matter remains close enough to the 
would-be-horizon, or if this process constitutes a transient phase after which the object relaxes to a stable state (maybe horizonless to avoid the self consistency issues of the other classes of geometries). Whether this is undesirable, or an opportunity for the observational detection of new physics, depends on philosophical considerations that are out of the scope of this present work. The most immediate consequence of our analysis, which is spelled out in more detail in the companion letter \cite{letter}, is that one cannot just naively assert that singularity regularization will solve the outstanding issues of black hole physics without introducing additional aspects that, after a thorough analysis, may lead to radical surprises for our theoretical and observational understanding of black holes.

Finally, let us note that, although we have assumed an everywhere well defined geometry, our analysis is valuable also when this is not the case and it allows us to understand how to construct singularity regularization scenario that solves the self consistency issues raised in this paper by allowing for region where an effective geometry cannot be defined. Interesting examples are constituted by the pictures studied in \cite{Ashtekar:2005cj,DeLorenzo2015,Bianchi:2018} where the geometry is defined almost everywhere but in a small region, a feature that allows these models to escape some of the self consistency issues raised here.

\section*{Acknowledgements}
\noindent
The authors would like to thank Carlos Barcel\'o and Carlo Rovelli for useful discussions.
MV was supported by the Marsden Fund, which is administered by the Royal Society of New Zealand. 
MV would like to thank SISSA and INFN (Trieste) for hospitality during the early phase of this work.


\appendix
\section{Regularity conditions for vanishing radius \label{sec:regconds}}

One of the properties that we are demanding to the spacetime geometry is the absence of divergences in curvature invariants. We devote this section to the extraction of the consequences of this demand. We will express these consequences in a form that is suitable for straightforward use in our main discussion. Let us focus our attention on spacetimes in which $r$ can vanish, thus defining a special region in spacetime; absence of curvature singularities will imply certain constraints on the metric functions around this region. In the static case, this was discussed for instance in \cite{Carballo-Rubio2018}.

Let us start assuming that the metric coefficients are either real analytic or finite (which is a weaker assumption) in this open set. Given that the exponential function is real analytic, both functions $F(v,r)$ and $\phi(v,r)$ must be either analytic or finite as well. It follows then that divergences in curvature invariants can be avoided if and only if, for $r=\epsilon\ll1$,
\begin{equation}\label{eq:regcond}
F(v,\epsilon)=1+\mathcal{O}(\epsilon^2), \qquad\text{and}\qquad \phi(v,\epsilon)=\Phi_0(v)+\mathcal{O}(\epsilon^2).
\end{equation}
It is worth noticing that this is the straightforward extension of the condition that should applies in the static case \cite{Carballo-Rubio2018}. To reach this conclusion, let us start with the following relation satisfied by the Ricci scalar $g^{ab}R_{ab}$:
\begin{equation}
\lim_{r\rightarrow0}r^2g^{ab}R_{ab}=2-2F(v,0).
\end{equation}
Hence, the condition $F(v,0)=1+\mathcal{O}(r^2)$ is necessary to avoid a quadratic divergence. But we are still left with a possible linear divergence, as 
\begin{equation}
\lim_{r\rightarrow0}r\,g^{ab}R_{ab}=4\left.\partial_r\phi(v,r)\right|_{r=0}.
\end{equation}
It is from this equation that the second condition in Eq. \eqref{eq:regcond} arises.
It is easy to check that if conditions \eqref{eq:regcond} are fulfilled, also the curvature invariants $R_{ab}R^{ab}$, $R_{abcd}R^{abcd}$ and $C_{abcd} C^{abcd}$ are finite.

Eq. \eqref{eq:regcond} is to be compared with the value $F(v,0)=-\infty$ for standard Schwarzschild and Reissner--Nordstr\"om black holes. The different value of $F(v,0)$ translates also into a very different quasi-local structure around $r=0$, namely that of a future inner trapping horizon or, in other words, a future inner trapped surface for every constant value of $v$. Indeed, asymptotic flatness and the requirement of existence of a future outer trapping horizon, together with Eq. \eqref{eq:regcond}, implies that there must be (at least) two radial points in which $F(v,r)$ vanishes for every constant value of $v$. To see this let us specialize the previously introduced null vectors $\{{\bm k},{\bm l } \}$ to the radial null vector fields
\begin{equation}\label{eq:kldef}
\bm{k}=-e^{\phi(v,r)}\partial_r\,,\qquad \bm{l}=\partial_v+\frac{1}{2}e^{-\phi(v,r)}F(v,r)\partial_r,
\end{equation}
which satisfy the normalization condition $\bm{k}\cdot\bm{l}=-1$. These two linearly independent radial null vector fields correspond to ingoing and outgoing radial null geodesics, respectively. The expansion along outgoing radial null geodesics is then given by
\begin{equation}\label{eq:outexp}
\theta^{(\bm{l})}(v,r)=\frac{e^{-\phi(v,r)}}{r}F(v,r).
\end{equation}
Asymptotic flatness implies that the expansion is positive in the limit $r\rightarrow\infty$. On the other hand, the existence of a future outer marginally trapped surface for each value of $v$ implies that there must exist a function $r_{\rm outer}(v)$ such that $F(v,r_{\rm outer}(v))=0$ vanishes. Unless the corresponding future outer trapping horizon is extremal, the function $F(v,r)$ will then change sign around $r=r_{\rm outer}(v)$, being positive just above this value of the radius, and negative just below. It is at this point in the discussion where Eq. \eqref{eq:regcond} enters: given that $F(v,0)=1>0$, there must exist another function $r_{\rm innter}(v)$ such that $F(v,r_{\rm inner}(v))$. It is straightforward to show that $r=r_{\rm outer}(v)$ corresponds to a future outer marginally trapped surface, while $r=r_{\rm inner}(v)$ to a future inner marginally trapped surface; the distinction between outer and inner is provided by the different signs of $\mathcal{L}_{\bm{k}}\theta^{(\bm{l})}$. In fact, from the definitions in Eq. \eqref{eq:kldef} follows that
\begin{align}\label{eq:expansions}
&\left.\mathcal{L}_{\bm{k}}\theta^{(\bm{l})}(v,r)\right|_{r=r_{\rm outer}(v)}=-\frac{1}{r}\left.\partial_rF(v,r)\right|_{r=r_{\rm outer}(v)}<0,\nonumber\\
&\left.\mathcal{L}_{\bm{k}}\theta^{(\bm{l})}(v,r)\right|_{r=r_{\rm inner}(v)}=-\frac{1}{r}\left.\partial_rF(v,r)\right|_{r=r_{\rm inner}(v)}>0.
\end{align}

The discussion above relies on both real functions $F(v,r)$ and $\phi(v,r)$ being finite at $r=0$ (or real analytic, which is stronger). Let us briefly analyze the possibility that these functions are divergent. It is still possible to write down constraints (taking the form of differential equations) to be satisfied by the functions $F(v,r)$ and $\phi(v,r)$ so that each independent curvature invariant remains finite. However, it is not clear that all these constraints are compatible between them and, therefore, whether it is possible to engineer these divergent functions at $r=0$ to avoid curvature invariants blowing up. Moreover, it is straightforward to see that, if such a geometry exists, it must be characterized by a non-polynomial divergence for  $F(v,r)$ and $\phi(v,r)$ at $r=0$. 
In fact, assuming for $r=\epsilon\ll1$ that
\begin{equation}
F(v,\epsilon)=\epsilon^{-\alpha}f(v),\qquad\qquad e^{-\phi(v,\epsilon)}=\epsilon^{-\beta}\tilde{\phi}(v),
\end{equation}
with both $\alpha$ and $\beta$ integers greater than one, it follows that 
\begin{equation}
\left.g^{ab}R_{ab}\right|_{r=\epsilon}=\frac{2}{\epsilon^2}-\frac{\left(\alpha ^2-3\alpha+3 \alpha  \beta +2 \beta ^2-2\beta +2\right)f(v)}{\epsilon^{\alpha + 2}}.
\end{equation}
The polynomial $\alpha ^2-3\alpha+3 \alpha  \beta +2 \beta ^2-2\beta +2= \alpha^2+3(\beta-1)\alpha+\beta^2+1+(\beta-1)^2$ does not have roots in the domain of definition of $\alpha$ and $\beta$. Hence, any non-zero $\alpha$ leads to a curvature singularity. A curvature singularity can be avoided if and only if $\alpha=0$ and
\begin{equation}\label{eq:curv_sing}
f(v)=\frac{1}{\beta^2-\beta+1}>0.
\end{equation}
This would already imply the existence of an inner horizon in complete parallel to our previous discussion. However, computing the Kretschmann scalar shows that
\begin{equation}
\lim_{r\rightarrow0}r^4 R_{abcd}R^{abcd}=\frac{8 \beta ^2 \left(\beta ^2+2\right)}{\left(\beta ^2-\beta +1\right)^2 }
\end{equation}
This implies the necessity of the further condition $\beta=0$.  Hence, even if allowing that $F(v,r)$ and $\phi(v,r)$ diverge polynomially at $r=0$, in order to avoid that some curvature invariant blows up it is necessary to assume that $F(v,0)=1$.

\bibliography{refs}
\end{document}